\documentclass[ALICE,manyauthors]{cernphprep}
\usepackage[comma,square,numbers,sort&compress]{natbib}
\usepackage{hyperref}
\usepackage{lineno}
\usepackage[T1]{fontenc}
\usepackage{xspace}
\usepackage[utf8]{inputenc}
\usepackage{booktabs}
\usepackage{multirow}
\usepackage{textcomp}
\usepackage[dvipsnames]{xcolor}
\usepackage{orcidlink}

\begin{document}
%

\newcommand{\pp}           {pp\xspace}
\newcommand{\ppbar}        {\mbox{$\mathrm {p\overline{p}}$}\xspace}
\newcommand{\XeXe}         {\mbox{Xe--Xe}\xspace}
\newcommand{\PbPb}         {\mbox{Pb--Pb}\xspace}
\newcommand{\pA}           {\mbox{pA}\xspace}
\newcommand{\pPb}          {\mbox{p--Pb}\xspace}
\newcommand{\AuAu}         {\mbox{Au--Au}\xspace}
\newcommand{\dAu}          {\mbox{d--Au}\xspace}

\newcommand{\s}            {\ensuremath{\sqrt{s}}\xspace}
\newcommand{\snn}          {\ensuremath{\sqrt{s_{\mathrm{NN}}}}\xspace}
\newcommand{\pt}           {\ensuremath{p_{\rm T}}\xspace}
\newcommand{\meanpt}       {$\langle p_{\mathrm{T}}\rangle$\xspace}
\newcommand{\ycms}         {\ensuremath{y_{\rm CMS}}\xspace}
\newcommand{\ylab}         {\ensuremath{y_{\rm lab}}\xspace}
\newcommand{\etarange}[1]  {\mbox{$\left | \eta \right |~<~#1$}}
\newcommand{\yrange}[1]    {\mbox{$\left|y\right|<#1$}}
\newcommand{\dndy}         {\ensuremath{\mathrm{d}N_\mathrm{ch}/\mathrm{d}y}\xspace}
\newcommand{\dndeta}       {\ensuremath{\mathrm{d}N_\mathrm{ch}/\mathrm{d}\eta}\xspace}
\newcommand{\avdndeta}     {\ensuremath{\langle\dndeta\rangle}\xspace}
\newcommand{\dNdy}         {\ensuremath{\mathrm{d}N_\mathrm{ch}/\mathrm{d}y}\xspace}
\newcommand{\Npart}        {\ensuremath{N_\mathrm{part}}\xspace}
\newcommand{\Ncoll}        {\ensuremath{N_\mathrm{coll}}\xspace}
\newcommand{\dEdx}         {\ensuremath{\textrm{d}E/\textrm{d}x}\xspace}
\newcommand{\RpPb}         {\ensuremath{R_{\rm pPb}}\xspace}
\newcommand{\avpt}         {\ensuremath{\langle\pt\rangle}\xspace}
\newcommand{\multdensmid}  {\ensuremath{\langle\dndeta\rangle_{|\eta|~<~0.5}}\xspace}
\newcommand{\inelgz}       {\ensuremath{{\rm INEL>0}}\xspace}

\newcommand{\nineH}        {$\sqrt{s}~=~0.9$~Te\kern-.1emV\xspace}
\newcommand{\seven}        {$\sqrt{s}~=~7$~Te\kern-.1emV\xspace}
\newcommand{\twoH}         {$\sqrt{s}~=~0.2$~Te\kern-.1emV\xspace}
\newcommand{\twosevensix}  {$\sqrt{s}~=~2.76$~Te\kern-.1emV\xspace}
\newcommand{\five}         {$\sqrt{s}~=~5.02$~Te\kern-.1emV\xspace}
\newcommand{\thirteen}         {$\sqrt{s}~=~13$~Te\kern-.1emV\xspace}
\newcommand{\twosevensixnn}{$\sqrt{s_{\mathrm{NN}}}~=~2.76$~Te\kern-.1emV\xspace}
\newcommand{\fivenn}       {$\sqrt{s_{\mathrm{NN}}}~=~5.02$~Te\kern-.1emV\xspace}
\newcommand{\LT}           {L{\'e}vy-Tsallis\xspace}
\newcommand{\GeVc}         {Ge\kern-.1emV/$c$\xspace}
\newcommand{\MeVc}         {Me\kern-.1emV/$c$\xspace}
\newcommand{\TeV}          {Te\kern-.1emV\xspace}
\newcommand{\GeV}          {Ge\kern-.1emV\xspace}
\newcommand{\MeV}          {Me\kern-.1emV\xspace}
\newcommand{\GeVmass}      {Ge\kern-.1emV/$c^2$\xspace}
\newcommand{\MeVmass}      {Me\kern-.1emV/$c^2$\xspace}
\newcommand{\lumi}         {\ensuremath{\mathcal{L}}\xspace}

\newcommand{\ITS}          {\rm{ITS}\xspace}
\newcommand{\TOF}          {\rm{TOF}\xspace}
\newcommand{\ZDC}          {\rm{ZDC}\xspace}
\newcommand{\ZDCs}         {\rm{ZDCs}\xspace}
\newcommand{\ZNA}          {\rm{ZNA}\xspace}
\newcommand{\ZNC}          {\rm{ZNC}\xspace}
\newcommand{\SPD}          {\rm{SPD}\xspace}
\newcommand{\SDD}          {\rm{SDD}\xspace}
\newcommand{\SSD}          {\rm{SSD}\xspace}
\newcommand{\TPC}          {\rm{TPC}\xspace}
\newcommand{\TRD}          {\rm{TRD}\xspace}
\newcommand{\VZERO}        {\rm{V0}\xspace}
\newcommand{\VZEROA}       {\rm{V0A}\xspace}
\newcommand{\VZEROC}       {\rm{V0C}\xspace}
\newcommand{\Vdecay} 	   {\ensuremath{V^{0}}\xspace}

\newcommand{\ee}           {\ensuremath{e^{+}e^{-}}} 
\newcommand{\pip}          {\ensuremath{\pi^{+}}\xspace}
\newcommand{\pim}          {\ensuremath{\pi^{-}}\xspace}
\newcommand{\kap}          {\ensuremath{\rm{K}^{+}}\xspace}
\newcommand{\kam}          {\ensuremath{\rm{K}^{-}}\xspace}
\newcommand{\pbar}         {\ensuremath{\rm\overline{p}}\xspace}
\newcommand{\kzero}        {\ensuremath{{\rm K}^{0}_{\rm{S}}}\xspace}
\newcommand{\lmb}          {\ensuremath{\Lambda}\xspace}
\newcommand{\almb}         {\ensuremath{\overline{\Lambda}}\xspace}
\newcommand{\Om}           {\ensuremath{\Omega^-}\xspace}
\newcommand{\Mo}           {\ensuremath{\overline{\Omega}^+}\xspace}
\newcommand{\X}            {\ensuremath{\Xi^-}\xspace}
\newcommand{\Ix}           {\ensuremath{\overline{\Xi}^+}\xspace}
\newcommand{\Xis}          {\ensuremath{\Xi^{\pm}}\xspace}
\newcommand{\Oms}          {\ensuremath{\Omega^{\pm}}\xspace}
\newcommand{\degree}       {\ensuremath{^{\rm o}}\xspace}
\newcommand{\Xiboth}       {\ensuremath{\Xi}\xspace}
\newcommand{\Omboth}       {\ensuremath{\Omega}\xspace}
\newcommand{\vzeros}       {\ensuremath{\rm{V}^{0}}\xspace}

\begin{titlepage}
\PHyear{2025}       
\PHnumber{256}      
\PHdate{05 November}  

\title{Strangeness production as a function of charged-particle multiplicity in proton--proton collisions at ${\bf \sqrt{s}~=~5.02}$~TeV}
\ShortTitle{Strangeness production in pp collisions at \five}

\Collaboration{ALICE Collaboration\thanks{See Appendix~\ref{app:collab} for the list of collaboration members}}
\ShortAuthor{ALICE Collaboration} 

\begin{abstract}
(Multi-)strange particle production rates and transverse momentum distributions are measured at midrapidity (\yrange0.5) as a function of the charged-particle multiplicity density by the ALICE experiment at the LHC, using proton--proton collisions at a center-of-mass energy of \five. 
This study extends similar studies performed at \seven and \thirteen to a lower energy regime, improving the statistical precision and extending the measurement to previously unexplored low-multiplicity regions. While \kzero, \lmb, and \Xiboth yields can be described with a linear multiplicity dependence within uncertainties, the \Omboth yields follow a significantly faster than linear increasing trend. For all analyzed particles, the overall production rate is consistent with those observed at higher energy and at similar multiplicity densities. Transverse momentum distributions are observed to evolve with multiplicity. Several state-of-the-art QCD-inspired Monte Carlo models have been compared to the data, testing some recently introduced features to address the findings at higher energies.
Models can qualitatively describe the transverse momentum spectra
and the \lmb/\kzero spectral ratio only if collectivity is introduced in the evolution of the system.
\end{abstract}
\end{titlepage}

\setcounter{page}{2} 


\section{Introduction} 
Strange hadron formation in high-energy hadronic collisions is the subject of an intense experimental study which aims at understanding the properties of quantum chromodynamics (QCD) in its non-perturbative regime.
Several QCD-inspired phenomenological Monte Carlo (MC) generators can be employed to predict the relative yields of different particle species in elementary hadronic collisions, but a comprehensive quantitative phenomenological description is still missing.

In heavy-ion collisions, an enhancement in the yield of strange hadrons relative to particles containing only u and d quarks was originally proposed as a signature of quark--gluon plasma (QGP) formation~\cite{Rafelski:1982, Rafelski:1980}. Strangeness enhancement has been experimentally observed over the last thirty years in heavy-ion collisions, first at the CERN SPS~\cite{Andersen:1999, Afanasiev:2002, Antinori:2010} and later at RHIC~\cite{Abelev:2008} and at the LHC~\cite{ALICE:2022wpn,ALICEstrPbPb276}. More recently, such trends have been interpreted, in the context of the Statistical Hadronization Model (SHM) description of particle production, as originating from the change in the source volume in different colliding systems~\cite{ANDRONIC2009142, WHEATON200984, VOVCHENKO2019295}. The ALICE collaboration recently reported on the charged-particle multiplicity dependence of strange and multi-strange particle production, showing a significant strangeness enhancement trend in proton--proton (\pp) and proton--lead (\pPb) collisions~\cite{ALICEstrNP, ALICEstrPP7, ALICEstrPP13}.
Strange particle yields are observed to be enhanced in high-multiplicity proton--proton collisions with respect to the corresponding yields measured in inclusive inelastic collisions at the same energy. In the highest multiplicity regime, relative particle abundances were found to reach values comparable to the ones observed in very central heavy-ion collisions. The progression with multiplicity of particle yields was observed to be independent of the collision energy, by comparing the \seven and \thirteen results ~\cite{ALICEstrPP13}. 
Moreover, transverse momentum spectra were observed to evolve towards higher average transverse momentum, \avpt, when the multiplicity density in the collision increases, a feature similar to what is observed in nucleus--nucleus interactions.

In order to interpret these observations, phenomenological models involving string fragmentation, typically employed in elementary leptonic and hadronic collisions~\cite{pythia8generic, HERWIGgeneric}, have been adapted by introducing final-state mechanisms which can enhance the number of baryons and strange particles in hadronic interactions by modifying the hadronization mechanisms in high-density environments~\cite{ColRecBeyondLead, HERWIGimproved,bierlich2015effects,PetroviciHijing}. At the same time, statistical models~\cite{FISTsmallsys, THERMUSsmallsys} and core-corona models~\cite{EPOScoco, DCCIcoco} have been exploited, extending their predictions from the high-multiplicity to the low-multiplicity regime.

In this paper, we report on the study of \kzero, \lmb, \almb, \X, \Ix, \Om and \Mo production, making use of the large data set ($\mathcal{L}_{\rm int} = 2.3~{\rm nb^{-1}}$) collected during the 2017 proton--proton data-taking campaign at \five. This is the largest data set analyzed for hadrochemistry studies in hadronic collisions, allowing for a significant improvement in the precision of the transverse momentum spectra of \Xis and \Oms baryons. The highest possible granularity in the multiplicity slicing was favored over a larger \pt reach, allowing the measurements to be extended down to the lowest multiplicities ever probed at the LHC.
In the following, particles and antiparticles are added together, as it has been demonstrated that their production yields are compatible at the LHC~\cite{ALICEstrNP} and are hereafter referred to as \lmb, \Xiboth and \Omboth, respectively.

In Sec.~2, the experimental apparatus used for the measurement are discussed, while in Sec.~3 and 4 the analysis technique and the evaluation of the systematic uncertainties are described, respectively. The results are presented in Sec.~5, and they are compared in Sec.~6 to phenomenological models. Finally, conclusions are drawn in Sec.~7.

\section{Experimental Setup}
Details on the ALICE detectors and their performance are described in Refs.~\cite{Aamodt:2008,Abelev:2014}. The detectors used for the analysis presented in this paper are the Inner Tracking System (ITS) ~\cite{Aamodt:2010}, the Time Projection Chamber (TPC) ~\cite{Alme:2010}, the Time-of-Flight (TOF) detector~\cite{Akindinov:2013}, and the two scintillator arrays of the V0 detector~\cite{Abbas:2013}. The ITS, TPC and TOF are part of the central barrel and are cylindrically arranged around the interaction point (IP), enclosed within the L3 solenoid magnet, which provides a field up to 0.5 T. Unless specified otherwise, the pseudorapidity coverage of the detectors in the central barrel is $|\eta| <0.9$. The ITS is the detector closest to the beam pipe and the IP. It comprises six cylindrical layers of silicon micro-pattern detectors with a two-dimensional readout. The innermost layers, for which a higher granularity
is required, are made of Silicon Pixel Detectors (SPD). The next two layers are covered by Silicon Drift Detectors (SDD), while the outer layers consist of Silicon Strip Detectors (SSD). 
The TPC is the main tracking detector in ALICE. It is a cylindrical gaseous detector with an internal volume of 88 m$^{3}$ surrounding the ITS. Its timing and spatial information provides up to 159 three-dimensional space points as well as the specific energy loss of the particles of interest. The TPC also provides particle identification through a d$E$/d$x$ measurement which covers the low-momentum region up to a few tens of GeV/$c$, with a resolution around 5-7$\%$. The outermost detector used is the TOF, positioned at a distance of 370~cm. It has full azimuthal acceptance, and an overall time resolution down to $\sim60$ ps, and is used to measure the time of flight.  The V0 detector is made up of two scintillator hodoscopes located around the beam pipe on opposite sides of the IP at $z$ = +340 cm (V0A) and $z$ = --90 cm (V0C) and covers $2.8 <\eta<5.1$ and $-3.7<\eta<-1.7$, respectively. It provides the minimum-bias trigger and the identification and rejection of beam-gas events. Additionally, the V0A and V0C detectors (together defined as V0M) are used to determine the charged-particle multiplicity density. 

\section{Data Analysis}
\label{sec:analysis}
The analysis presented in this paper follows a strategy similar to the one described in previous ALICE publications ~\cite{ALICEstrPP13, ALICEstrNP, ALICEstrPP7, ALICEstrPP7MB, ALICEstrPP09}. 
Primary strange hadrons were measured at central rapidity (\yrange0.5) in events with at least one charged particle produced in the pseudorapidity interval \etarange1 (\inelgz), corresponding to about $75 \%$ of the total inelastic cross section~\cite{ALICEmultPPmany}. Events were required to have a reconstructed vertex within the longitudinal interval $|z|<10$ cm from the IP, while events containing more than one distinct vertex were tagged as pileup and discarded.
In order to study the multiplicity dependence of strange hadron production, the sample was divided into event classes based on the total charge deposited in both the V0 detectors (V0M amplitude), which is proportional to the total number of charged particles detected by the V0 scintillators. The event classes are summarized in Table~\ref{tab:V0Mclass}. For each class, the average primary charged-particle density \multdensmid at midrapidity was calculated, fully corrected for acceptance, tracking and vertexing efficiency as well as for contamination from secondary particles and combinatorial background. Note that the midrapidity multiplicity distribution corresponding to each V0M class has a large relative root mean square (RMS), ranging from 70\% to 30\% for the lowest and highest multiplicity classes, respectively. More details can be found in Ref.~\cite{ALICEmultPPmany}.

\begin{table}[ht] 
   \caption{List of event multiplicity classes (first column), percentile intervals (second column) and corresponding charged-particle pseudorapidity densities (\etarange0.5) for pp collisions at \five in the INEL$>0$ event class~\cite{ALICEmultPPmany} for the different particles under study (\kzero and \lmb in third, \Xiboth in fourth and \Omboth in fifth column, respectively). The total systematic uncertainties are the quadratic sum of the correlated and uncorrelated systematic uncertainties.}
   \centering
   \begin{tabular}{|c||c||c|c|c|} 
   \hline
   \multirow{2}{*}{Class}   &  \multirow{2}{*}{Multiplicity (\%)}   & \multicolumn{3}{c|}{ $\avdndeta_{|\eta| < 0.5}$} \\
   \cline{3-5}
                            &                                       & \kzero, \lmb  & \Xiboth  & \Omboth \\
   \hline\hline     
   $\text{I}$   & $0-0.0091$    & $24.53 \pm 0.39$  & \multirow{2}{*}{$22.00 \pm 0.40$}    & \multirow{3}{*}{$18.50 \pm 0.17$}\\
   \cline{1-3}
   $\text{II}$  & $0.0091-0.0915$    & $21.71 \pm 0.21$  &                                       & \\
   \cline{1-4}
   $\text{III}$ & $0.0915-0.9152$       & $18.11 \pm 0.18$  & $18.11 \pm 0.18$                     &  \\
   \hline
   $\text{IV}$  & $0.9152-4.577$         & $14.51 \pm 0.18$  & $14.51 \pm 0.18$                     & $14.51 \pm 0.18$ \\
   \hline
   $\text{V}$   & $4.577-9.156$        & $11.93 \pm 0.11$  & $11.93 \pm 0.11$                     & \multirow{2}{*}{$11.24 \pm 0.13$}\\
   \cline{1-4}
   $\text{VI}$  & $9.156-13.74$       & \multirow{2}{*}{$9.70 \pm 0.13$} & $10.30 \pm 0.13$& \\
   \cline{1-2}
   \cline{4-5}
   $\text{VII}$ & $13.74-18.32$       &                                   & $9.12 \pm 0.12$ & \multirow{2}{*}{$8.38 \pm 0.12$} \\
   \cline{1-4}
   $\text{VIII}$& $18.32-27.51$       & $7.76 \pm 0.11$  & $7.76 \pm 0.11$ & \\
   \hline
   $\text{IX}$  & $27.51-36.76$       & $6.34 \pm 0.08$   & $6.34 \pm 0.08$ & \multirow{2}{*}{$5.78 \pm 0.08$}    \\
   \cline{1-4}
   $\text{X}$   & $36.76-46.11$       & $5.22 \pm 0.07$ &  $5.22 \pm 0.07$ & \\
   \hline
   $\text{XI}$  & $46.11-65.45$       & $3.94 \pm 0.06$  & $3.94 \pm 0.06$ & $3.94 \pm 0.06$\\
   \hline
   $\text{XII}$ & $65.45-100$      & $2.42 \pm 0.04$  & $2.42 \pm 0.04$ & $2.42 \pm 0.04$\\
   \hline
   \hline
   $\text{\inelgz}$  & $0-100$   & \multicolumn{3}{c|}{$5.48\pm 0.07$} \\ \hline
   \end{tabular}
   \label{tab:V0Mclass}
\end{table}

The strange hadrons \kzero and \lmb (\vzeros) and \Xiboth and \Omboth (cascades) were each reconstructed via their specific weak decay topology~\cite{pdg}:
\begin{align*}
    &\kzero  \rightarrow \pip + \pim                   & &\mbox{B.R.} = (69.20 \pm 0.05) \% \\ 
    &\lmb    \rightarrow \rm{p} + \pim     \mbox{(+ c.c.)}  & &\mbox{B.R.} = (63.9 \pm 0.5) \% \\
    &\X      \rightarrow \lmb + \pim  \mbox{(+ c.c.)}  & & \mbox{B.R.} = (99.887 \pm 0.035) \% \\ 
    &\Om     \rightarrow \lmb + \kam  \mbox{(+ c.c.)}  & & \mbox{B.R.} = (67.8 \pm 0.7) \%
\end{align*}
applying a series of track, candidate and topological selections as summarized in Table~\ref{tab:selections}. The selection criteria were previously discussed in detail in Ref.~\cite{ALICEstrPP13} and were optimized through specific studies performed on MC simulations.

In this analysis, ITS and TPC are the main detectors employed for reconstructing the trajectories of pions, protons and charged kaons, which constitute the tracks generated by the weak decay of the particles under study. The reconstructed tracks, selected in the pseudorapidity region \etarange0.8, were required to have a number of TPC crossed rows larger than 70 and a fraction of TPC crossed rows and findable clusters larger than 80\%.
Particle identification was performed by measuring the energy loss per unit of track length in the TPC gas volume and requiring it to be within $\pm4\sigma_{\rm{d}\it{E}/\rm{d}\it{x}}$ from the expected calculated average using the Bethe–Bloch parametrization, where $\sigma_{\rm{d}\it{E}/\rm{d}\it{x}}$ is the d$E$/d$x$ resolution.

Topological selections were applied to the reconstructed weak decay candidates to reduce the combinatorial background. The distance of closest approach (DCA) between the \vzeros decay tracks was required to be smaller than 1 (1.3) standard deviation in the \vzeros (cascades) analyses. The DCA between each \vzeros decay track and the primary collision vertex (PV) was required to be larger than 0.06 (0.12 or 0.13) cm for the \vzeros ($\Xi$ or $\Omega$) analyses. For the cascades, the DCA between the bachelor track (charged track from the first weak decay) and the primary vertex was required to be larger than 0.04 cm, the DCA between the \vzeros and the primary vertex was larger than 0.06 cm, while the DCA between the bachelor track and \vzeros was smaller than 1.3 cm. 
In the $\Xi$ analysis, an additional cut on the minimum DCA between the bachelor track and the baryon \vzeros decay track was applied (0.02 cm) to avoid the inclusion of candidates for which a decay product from the \vzeros meson was incorrectly assigned as the bachelor track. The cosine of the angle between the candidate momentum and the vector connecting the primary and secondary vertices (pointing angle) was required to be larger than 0.97 for each case except for the $\Lambda$, for which a stricter cut of 0.995 was applied.
The radial distance between the PV and the \vzeros vertex was set to be larger than 0.5 (1.2 or 1.1) cm for the \vzeros (\Xiboth or \Omboth) analyses, while in the cascade analysis, an additional requirement on the minimum transverse decay radius for the secondary vertex was defined at 0.6 (0.5) for $\Xi$ ($\Omega$). Reconstructed candidates were retained if the proper decay length was measured to be smaller than 20, 30, 15 and 7.5 cm for the \kzero, \lmb, \Xiboth and \Omboth analyses, respectively.
The \vzeros candidates from cascade decays were required to match an invariant mass window of $\pm$~8~\MeVmass with respect to the nominal \lmb mass. In the \Omboth analysis, an additional selection on the invariant mass starting from the alternative cascade hypothesis was implemented, requiring it to be separated by more than 8~\MeVmass from the nominal \Xiboth mass.
To reduce effects due to out-of-bunch pileup on the measured yields, it was required that at least one of the decay tracks either had a measured signal in the TOF detector or was reconstructed in both the ITS and TPC detectors.

\begin{table}[ht]
\caption{List of selection criteria applied to particle candidates and their decay tracks for the \kzero, \lmb, \Xiboth and \Omboth analyses.}
  \begin{center}
  \begin{tabular}{|l|c|c|}
    \hline
    \textbf{Topological Variable}  & \textbf{\kzero $(\Lambda)$} & \textbf{$\Xi$ $(\Omega)$} \\
    \hline
    DCA pos.(neg.) decay track to PV  & $>$ 0.06 \ cm & $>$ 0.12 (0.13) cm\\
    DCA \vzeros decay tracks  & $< 1.0 \ \sigma$  & $< 1.3 \sigma$\\
    DCA bachelor to PV & $-$ & $>$ 0.04 cm\\
    DCA \vzeros to PV  & $-$ & $>$ 0.06 cm\\
    DCA bachelor to \vzeros & $-$  & $<$ 1.3 cm\\
    DCA bachelor to baryon (only $\Xi$) & $-$ & $>$ 0.02 cm\\
    \vzeros cosine of pointing angle & $> 0.97 (0.995)$ & $>$ 0.97\\
    cascade cosine of pointing angle & $-$   & $> 0.97$\\
    \vzeros transverse decay radius  & $>$ 0.5 cm & $>$ 1.2 (1.1) cm\\
    cascade transverse decay radius & $-$ & $>$ 0.6 (0.5) cm\\
    \hline
    \hline
    \textbf{Candidate Selection}  & \textbf{\kzero $(\Lambda)$} & \textbf{$\Xi$ $(\Omega)$} \\
    \hline
    rapidity interval& $|y|<0.5$ & $|y|<0.5$\\
    proper decay length & $<$ 20 (30) cm & $<$ 15 (7.5) cm\\
    distance to competing cascade mass & $-$ & $-$ (8 MeV/$c^{2}$) \\
    \vzeros invariant mass window & $-$ & 8 MeV/$c^{2}$\\
    decay tracks with ITS refit or TOF matching &   $\ge1$   &     $\ge1$\\
    \hline
    \hline
    \textbf{Track Selection}  & \textbf{\kzero $(\Lambda)$} & \textbf{$\Xi$ $(\Omega)$}\\
    \hline
    TPC $\rm{d}\it{E}/\rm{d}\it{x}$ selection & $<4 \ \sigma$ & $<4 \ \sigma$\\
    pseudorapidity interval & $|\eta|<0.8$ & $|\eta|<0.8$\\
    TPC $N_{\rm{crossed}}/N_{\rm{findable}}$ &  $\ge 0.8$ & $\ge 0.8$\\
    TPC $N_{\rm{crossed~rows}}$ & $\geq70$ & $\geq70$\\
    \hline
  \end{tabular}
  \end{center}
\label{tab:selections}
\end{table}

Signal extraction from the invariant mass distribution of each candidate in \pt and multiplicity intervals was performed as previously detailed in Refs.~\cite{ALICEstrPP13, ALICEstrNP, ALICEstrPP7}. A preliminary fit of the invariant mass distribution was performed using a Gaussian for the signal peak plus a second-order polynomial function to describe the background. This allowed for extracting the mean ($\mu$) and width ($\sigma$) of the peak and modeling the background. The signal was then extracted with a bin counting procedure in a mass range within $\pm~5\sigma$ with respect to $\mu$, subtracting the integral of the background fit function in the same mass window. 

\lmb yields are affected by significant contamination from secondary particles, coming from the decay of charged and neutral \Xiboth. In order to estimate this contribution, the charged \Xiboth spectra (twice the particle/antiparticle spectrum to account for the neutral \Xiboth decay, assuming \X$^{\kern-.3em(+)}$/\Xiboth$^{\kern-.2em0}=1$) were folded with an MC-based \pt-binned 2D matrix describing the decay kinematics and incorporating secondary \lmb reconstruction efficiencies. Further details on MC simulations can be found below. The fraction of secondary \lmb particles in the measured spectrum varied between 10\% and 20\%, depending on \pt and multiplicity.

Raw \pt distributions were corrected for detector acceptance and efficiency, estimated using MC simulated events featuring the PYTHIA 8 generator ~\cite{pythia8generic} and GEANT 3 ~\cite{Brun:1993} transport code for the propagation through the detector material. Use of GEANT 4 ~\cite{Agostinelli:2003} was also tested, showing no significant difference compared to the default code.
The acceptance-times-efficiency was found to change with \pt, reaching values of about 40\%, 30\%, 30\%, and 20\% at intermediate \pt for \kzero, \lmb, \Xiboth, and \Omboth, respectively. These values include the branching ratio factors. They are observed to be independent of the multiplicity class to within 2\%, a value limited by the available MC simulated data. The dependence of the efficiency on the generated \pt distributions was checked for all particle species. This effect, only relevant when large \pt intervals are considered ($\Xi$ and $\Omega$ analyses), was removed by reweighting the MC \pt distribution in order to adapt it to the measured one using a rapidly-converging iterative procedure. The normalization factor for the \inelgz event class was corrected to account for events lost due to trigger inefficiency~\cite{ALICEmultPPmany}. Un-triggered events may also contain signal candidates, so that a \pt-dependent correction was estimated by exploiting the same MC simulations used for the evaluation of acceptance and efficiency.

In order to compute average \pt (\avpt) and the \pt-integrated production yields, the spectra were fitted with a Lévy-Tsallis~\cite{Tsallis:1988,Azmi:2014dwa} function to extrapolate the unmeasured \pt region down to \pt = 0. To avoid potential bias coming from the fit of large \pt bins in the cascades analysis and at high \pt, the integral of the function was fitted to the bin contents. The systematic uncertainties for this extrapolation procedure were evaluated using other fit functions, as discussed in the next section.

\section{Systematic Uncertainties}
Several sources of systematic uncertainties were considered, following the strategy previously reported in Ref.~\cite{ALICEstrPP13}. In order to avoid being limited by the MC sample size in the systematic uncertainty determination and since systematics were shown to be independent of multiplicity, the \inelgz class was used, and the same relative systematic uncertainty was assigned to all multiplicity classes.
The stability of the signal extraction method was tested by varying the width used to define the ``signal'' region, expressed in terms of the number of $\sigma$ (see Sec.~\ref{sec:analysis}), varying it from the default $\pm5\sigma$ to $\pm4\sigma$ and $\pm6\sigma$. This uncertainty, moderately \pt dependent, was below 1\% for \kzero and below 2\% for all other particles.
Acceptance and efficiency corrections can introduce a systematic bias if the track, candidate and topological selections differ in MC from real data. This uncertainty was estimated by varying all the selection criteria within ranges leading to a maximum variation of $\pm10\%$ in the raw signal yield. The results were compared to those obtained with the default selection criteria, considering a significant variation which is incompatible with statistical fluctuations following the prescription in Ref.~\cite{barlow} with a $1\sigma$ threshold. Significant variations, induced by each selection separately, were added in quadrature to evaluate the systematic uncertainty.
The resulting uncertainty from topological, candidate and track selections depends on \pt and amounts at most (at low \pt) to 1.5\%, 5\%, 6\%, and 4.5\% for \kzero, \lmb, \Xiboth, and \Omboth, respectively.
Furthermore, a 2\% uncertainty was added in quadrature to account for possible variations of the efficiency with multiplicity.
The uncertainty due to the approximate knowledge and description of the detector material was estimated with a MC simulation, where the material budget was varied within its uncertainty of $\pm4.5\%$~\cite{ALICEstrPP13}. The assigned systematic uncertainty ranges from 8\% to 1\% at low and high \pt for cascades, while it stays below 4\% for \kzero and \lmb.

While in-bunch pileup rejection turns out to have a negligible effect for the data-taking period under study (the probability of having more than one pp interaction in a bunch crossing is $\sim\kern-.3em10^{-3}$), out-of-bunch pileup rejection was necessary and implemented by requiring ITS or TOF matching (see Sec.~\ref{sec:analysis}). The systematic uncertainty for this rejection procedure was evaluated by requiring both (or at least two for cascades) daughter tracks to be reconstructed in the ITS or TOF. The difference from the default case was taken as the systematic uncertainty, which is found to increase with \pt and saturate at high \pt, reaching a maximum value of 3\%.

In the \lmb analysis, the feed-down correction method is a source of systematic uncertainty, due to the uncertainties of the measured \Xiboth spectrum and the assumption that the \lmb originates in equal amounts from neutral and charged \Xiboth. The measured cascade spectra were moved upwards and downwards by one $\sigma$ in the systematic uncertainty followed by a reevaluation of the feed-down. Moreover, the assumption of $\Xis/\Xi^{0} = 1$ was replaced by the ratio extracted from the same MC simulation used to evaluate the acceptance and efficiency correction. The total difference with respect to the standard approach ranges from 2\% to 5\% depending on \pt and multiplicity.
Systematics related to the normalization of the corrected spectra were evaluated using the EPOS-LHC MC instead of the default PYTHIA 8 simulations. The systematic errors obtained do not exceed 4\% at low multiplicity for all particles and they rapidly become negligible as the multiplicity increases. 

The \meanpt and integrated yields for all particles except for the \kzero were obtained from the extrapolation to the unmeasured region. This procedure leads to an additional systematic uncertainty due to the unknown functional form used to reproduce the extended \pt spectrum. As illustrated in Sec.~\ref{sec:analysis}, a default extrapolation was performed using the Lévy-Tsallis formula, which describes the \pt spectra for all the examined strange hadrons over the full measured \pt range.
The uncertainty for the extrapolated fraction was estimated by fitting the spectra with four alternative functions (Blast-Wave, Boltzmann, $m_{\rm T}$-exponential, Fermi--Dirac).
The resulting uncertainty for the integrated yields largely depends on multiplicity, ranging from 7\% at low multiplicity to 2.5\% at high multiplicity for \lmb, from 13\% to 1\% for \Xiboth and from 17\% to 3\% for \Omboth. The extrapolation uncertainty for \meanpt is generally smaller, approximately half of that for the yields.

\section{Results}
\begin{figure}[]
\centering
\includegraphics[scale=0.43]{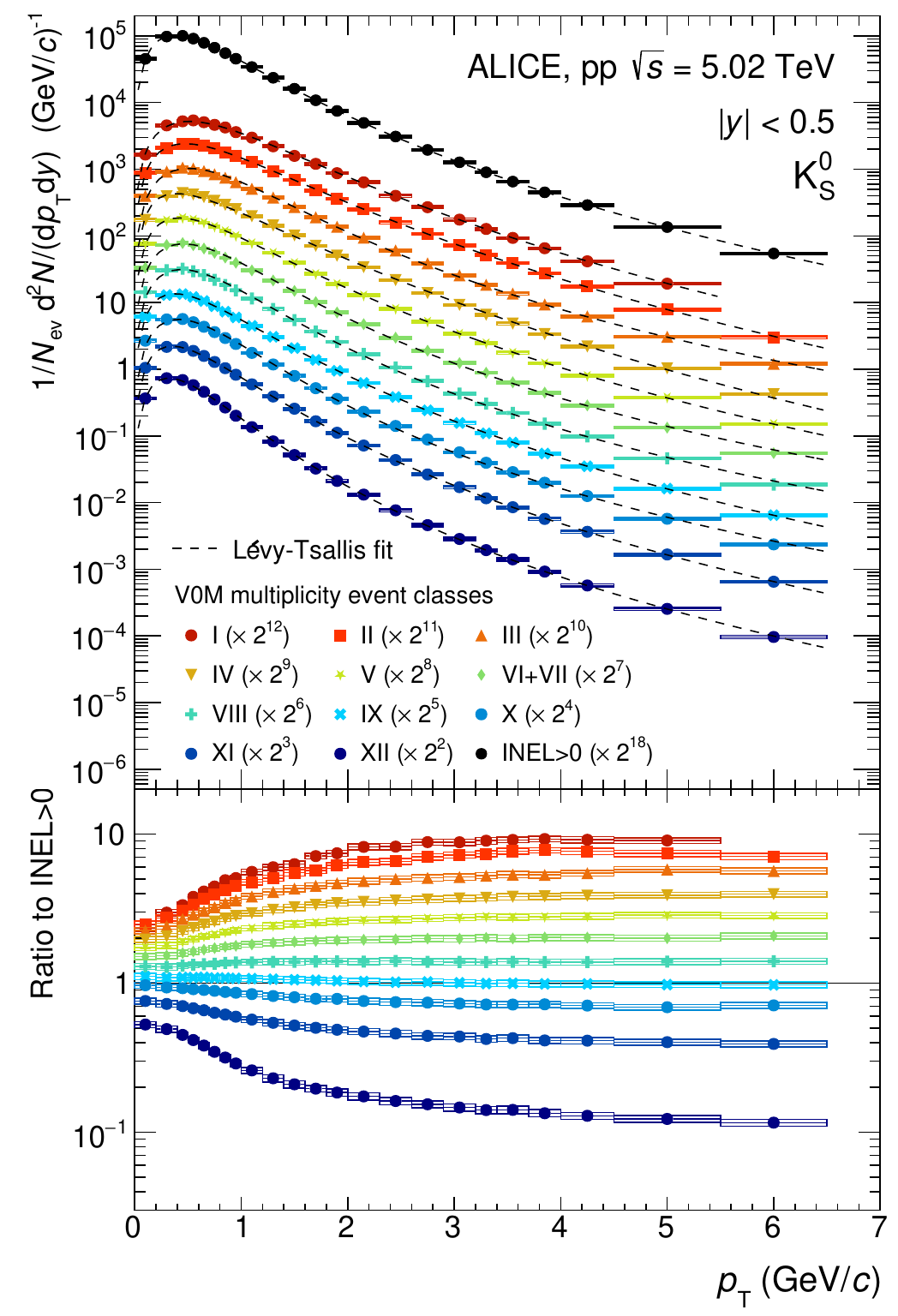}
\includegraphics[scale=0.43]{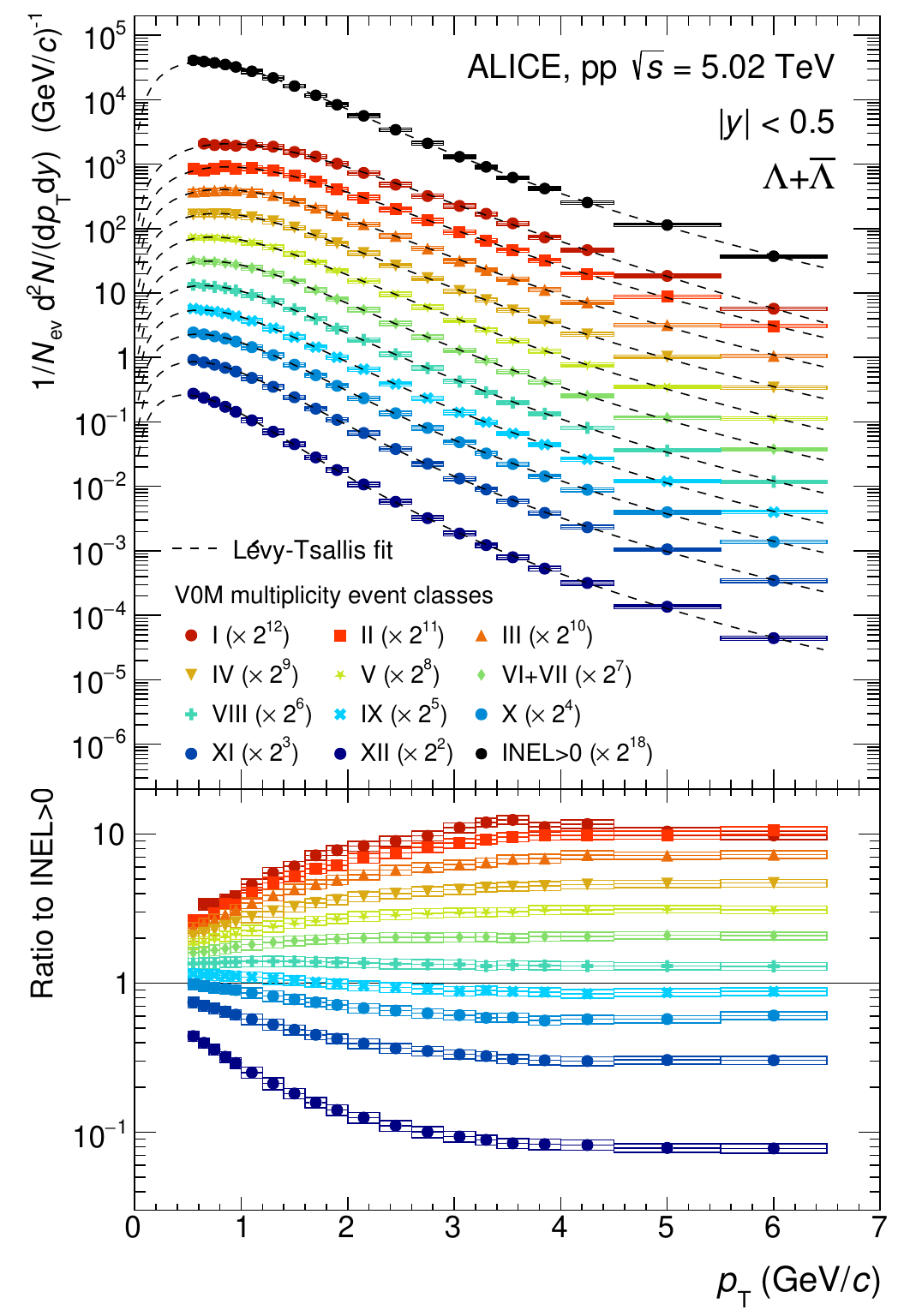}
\includegraphics[scale=0.43]{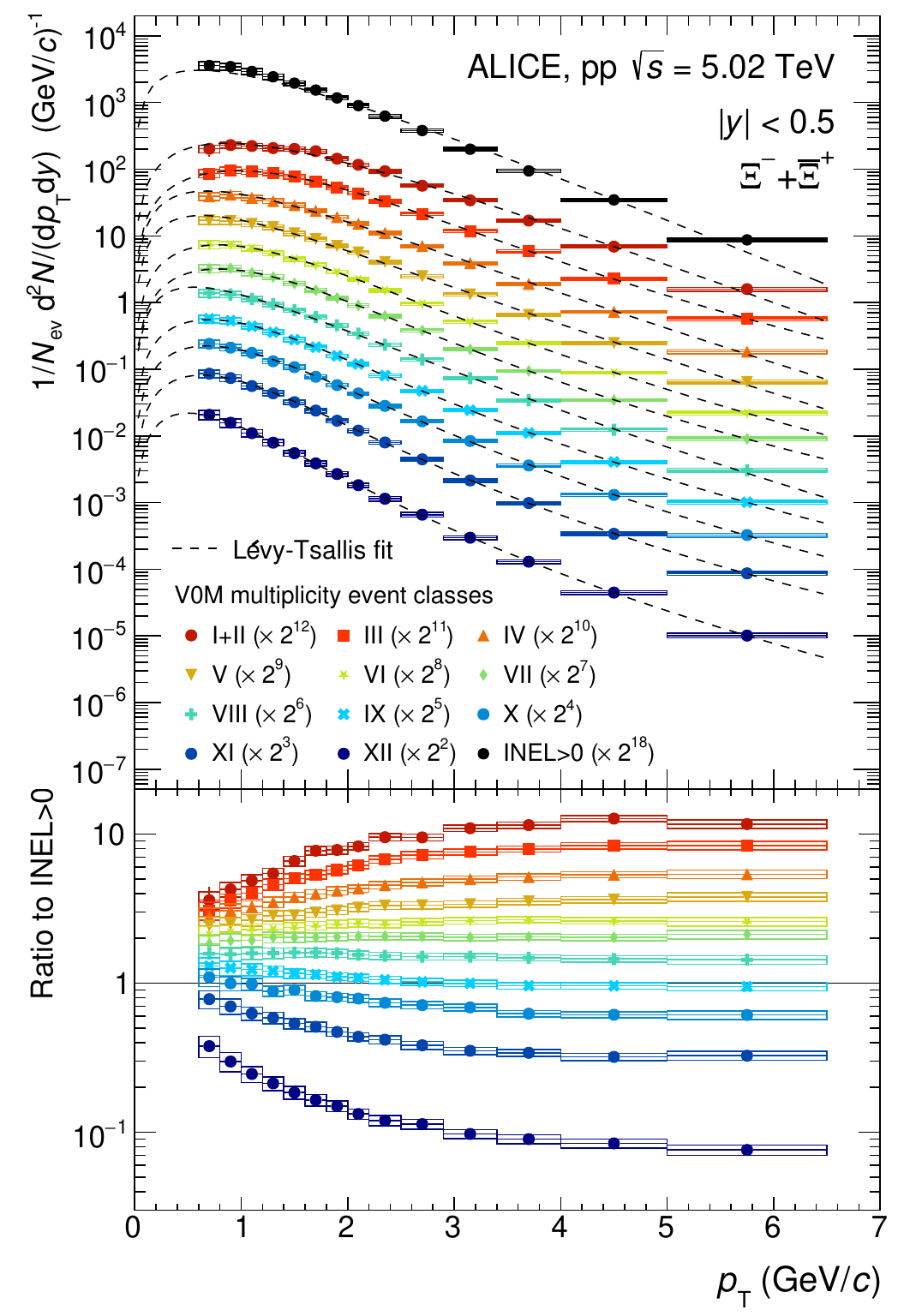}
\includegraphics[scale=0.43]{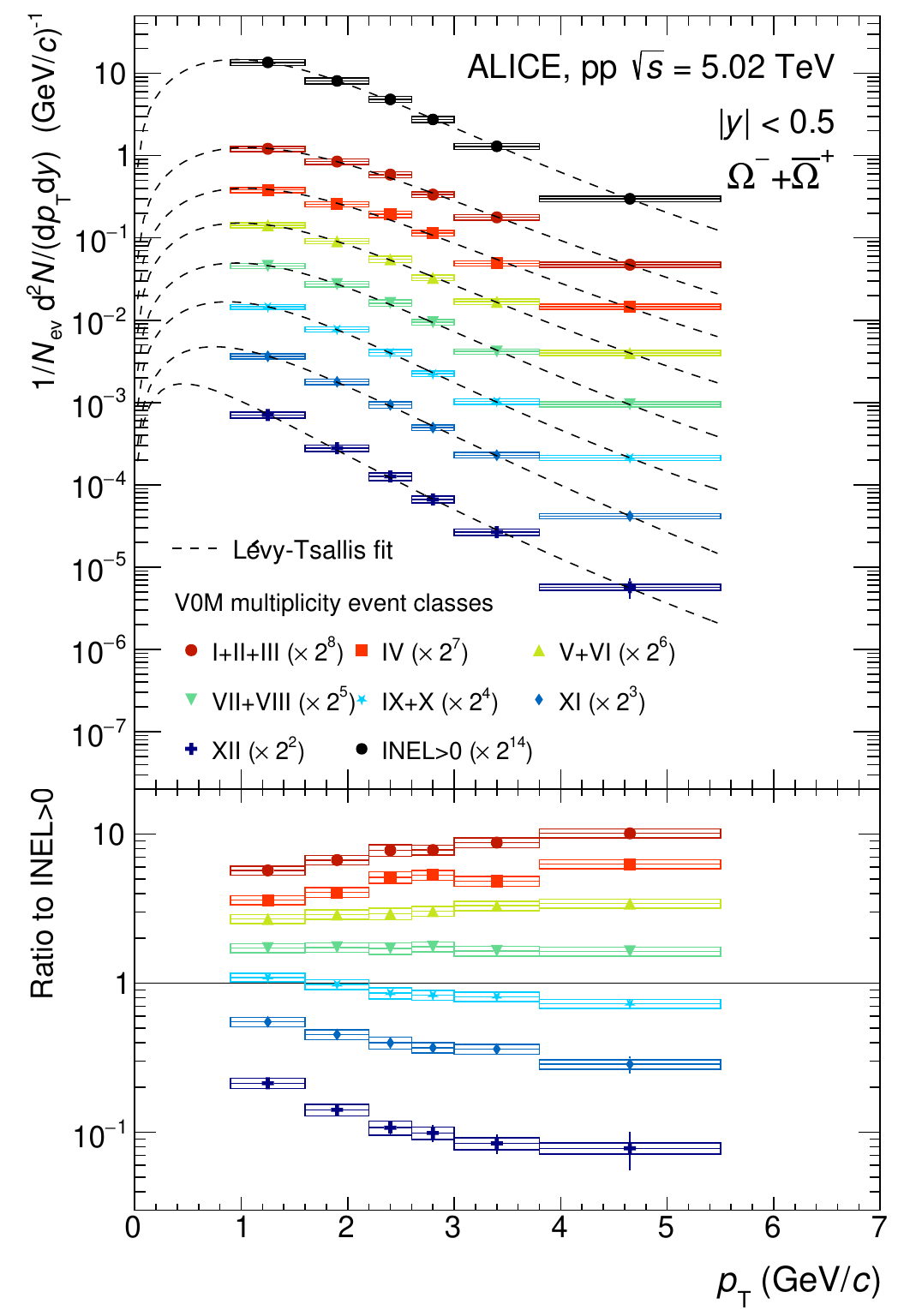}
\caption{\pt distributions at midrapidity for \kzero, \lmb, \Xiboth, and \Omboth for multiplicity classes and \inelgz, as indicated. Statistical uncertainties are shown as error bars, the systematic ones as boxes. The bottom panels present the ratio of all spectra to the \inelgz one. The dashed curves show the Lévy-Tsallis ~\cite{Tsallis:1988} fit to the \pt distributions.}
\label{fig:ptspectra}
\end{figure}

The \pt distributions of \kzero, \lmb, \Xiboth, and \Omboth for each multiplicity class and the multiplicity-integrated case are shown in Fig.~\ref{fig:ptspectra} together with the fit performed with a Lévy-Tsallis distribution. 
The \pt spectra feature a maximum which moves towards larger values as the multiplicity increases, a feature previously observed in large colliding systems and interpreted as the result of a collective boost in the radial direction. This feature is highlighted in the lower panels of Fig.~\ref{fig:ptspectra}, where the ratio of the multiplicity-differential spectrum to the \inelgz is reported for all particles under study. These ratios evolve significantly at low \pt, approaching a plateau at around 4~\GeVc, where they saturate for \kzero and \lmb. 
For the first time in pp collisions, the precision achieved with the present measurement allows such a saturation region to be approached for multi-strange hadrons as well, though measurements at even higher \pt would be necessary to strengthen the observation.

\begin{figure}[h]
\centering
\includegraphics[scale=0.39]{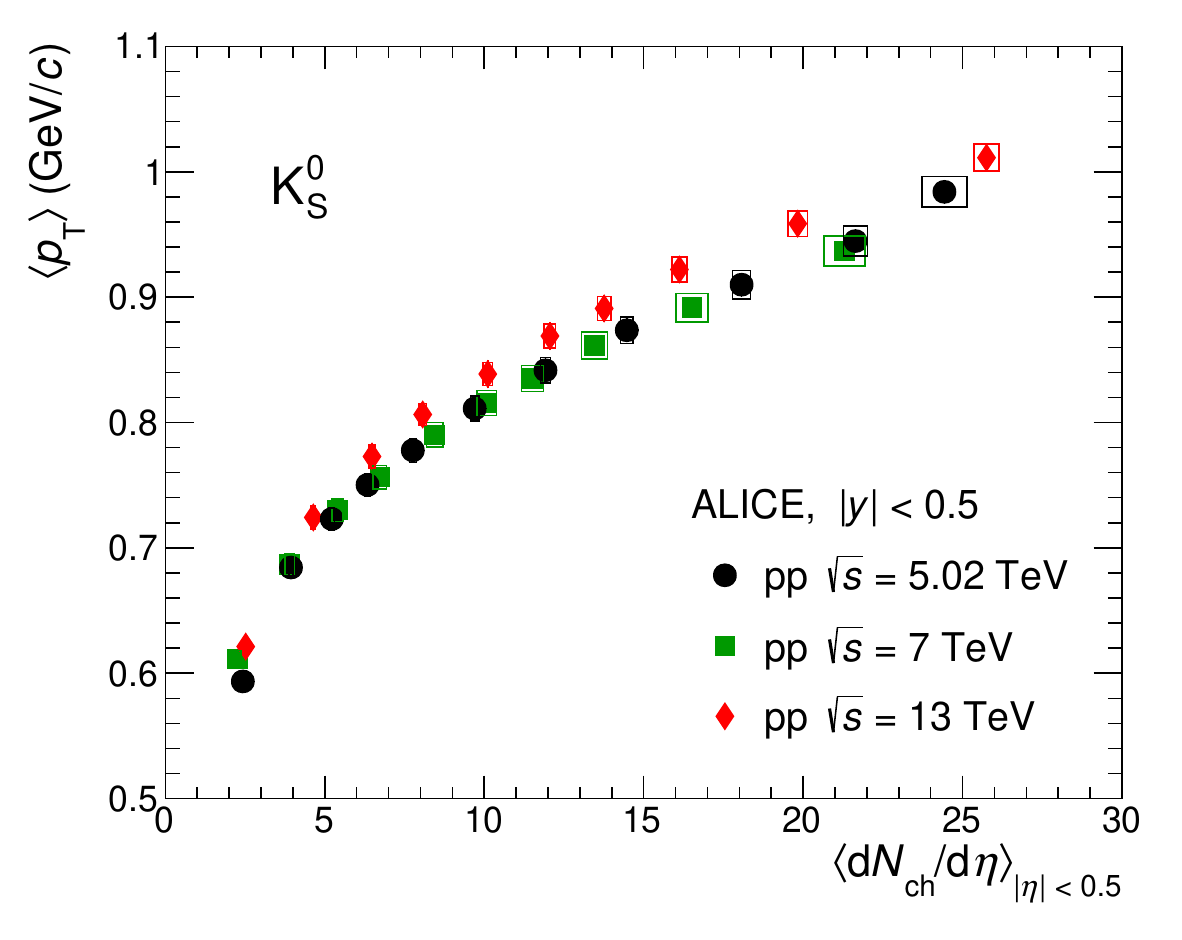}
\includegraphics[scale=0.39]{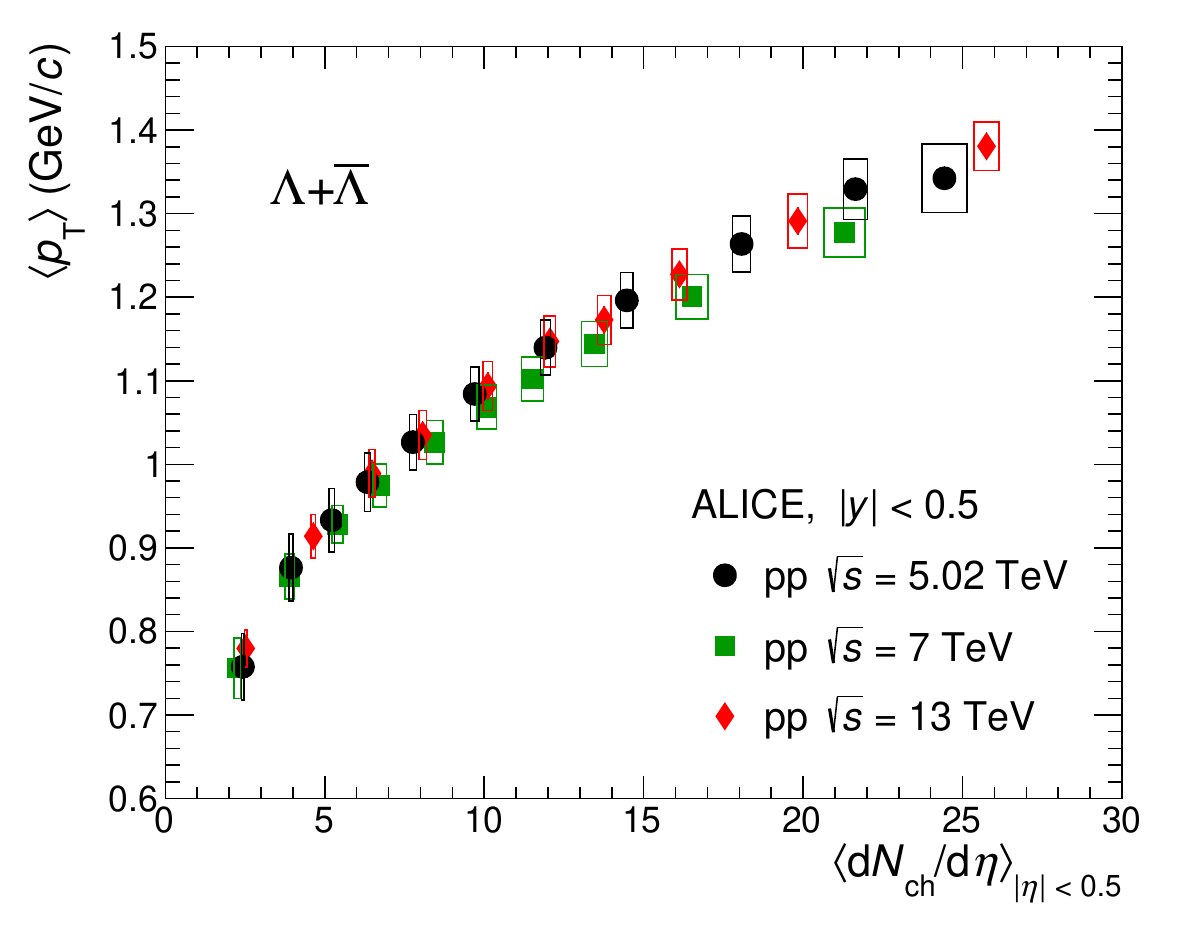}
\includegraphics[scale=0.39]{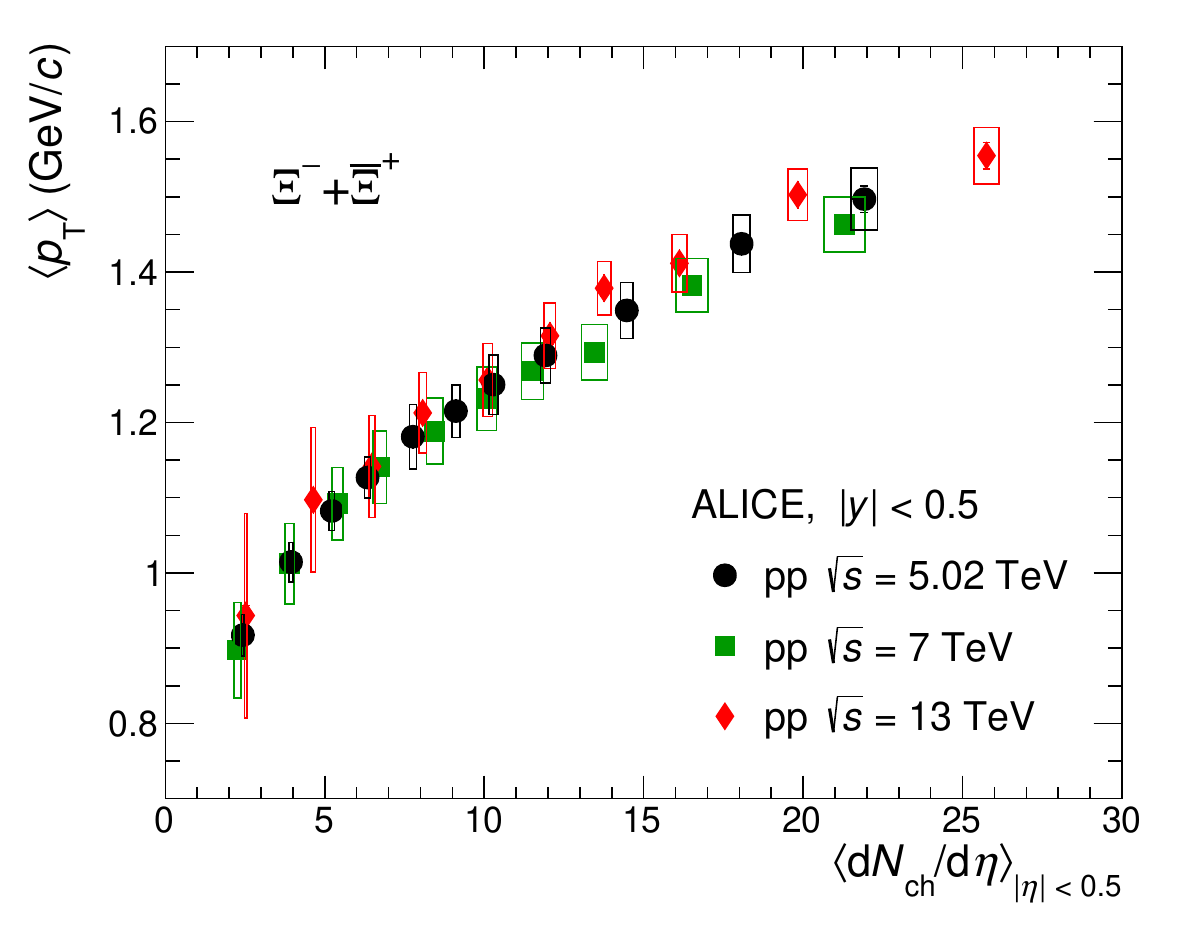}
\includegraphics[scale=0.39]{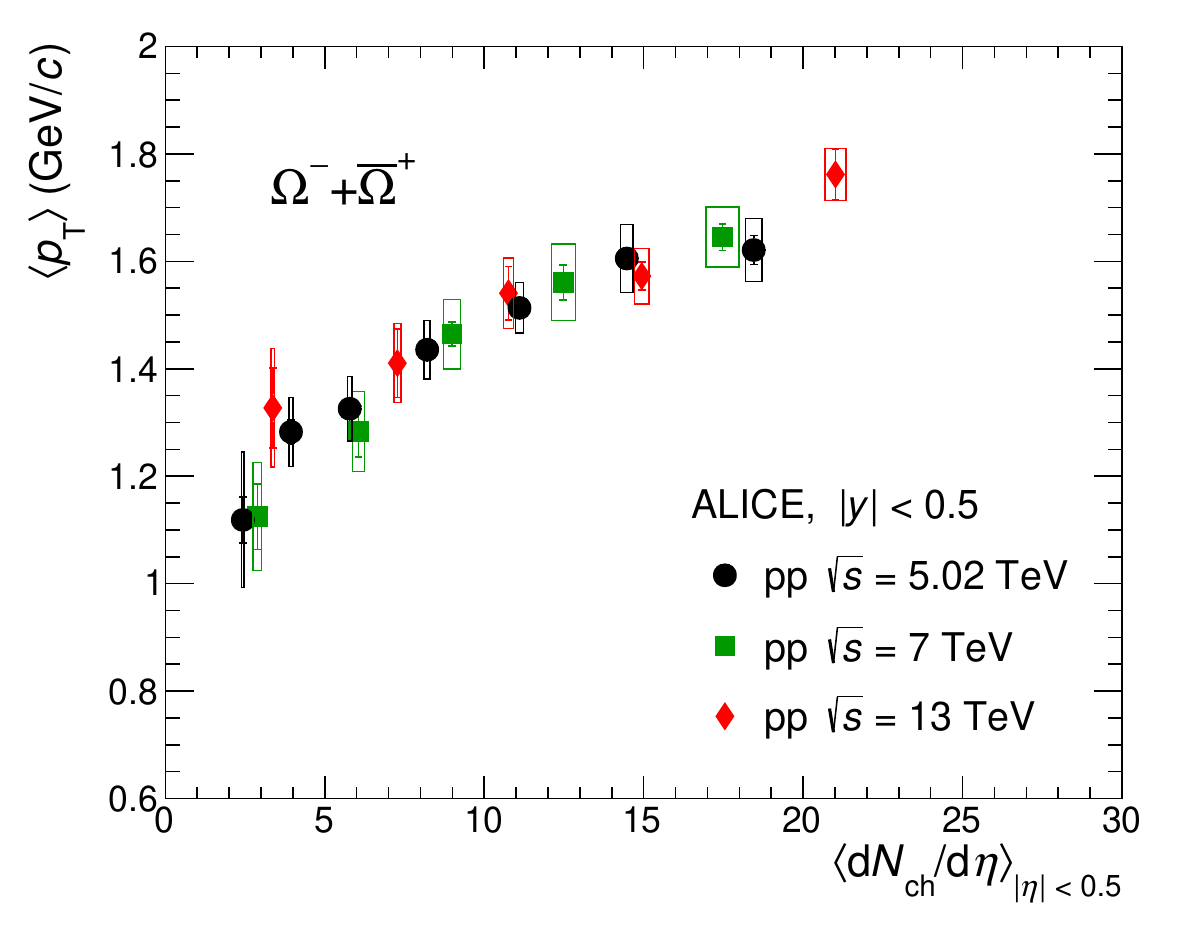}
\caption{\avpt in pp collisions at different center-of-mass energies and midrapidity as a function of the event multiplicity for \kzero and \lmb (top panels), \Xiboth and \Omboth (bottom panels). \five results are shown as black dots, \seven results~\cite{ALICEstrPP7} as green squares and \thirteen results \cite{ALICEstrPP13} as red diamonds. Statistical uncertainties are reported as error bars, systematic uncertainties as boxes whose width represent the systematic error on the multiplicity determination.}
\label{fig:meanpt}
\end{figure}

In order to quantify the evolution of the low-\pt region of the spectra with multiplicity, the \meanpt is presented in Fig.~\ref{fig:meanpt}. These values are compared to those obtained at higher energies in pp collisions by the ALICE Collaboration \cite{ALICEstrNP, ALICEstrPP13}. The increasing trend of the \meanpt is consistent for all center-of-mass energies, except for a 1$\sigma$ difference for the \kzero, for which results at the highest energy tend to be higher over the full multiplicity range.

\begin{figure}[]
\centering
\includegraphics[scale=0.44]{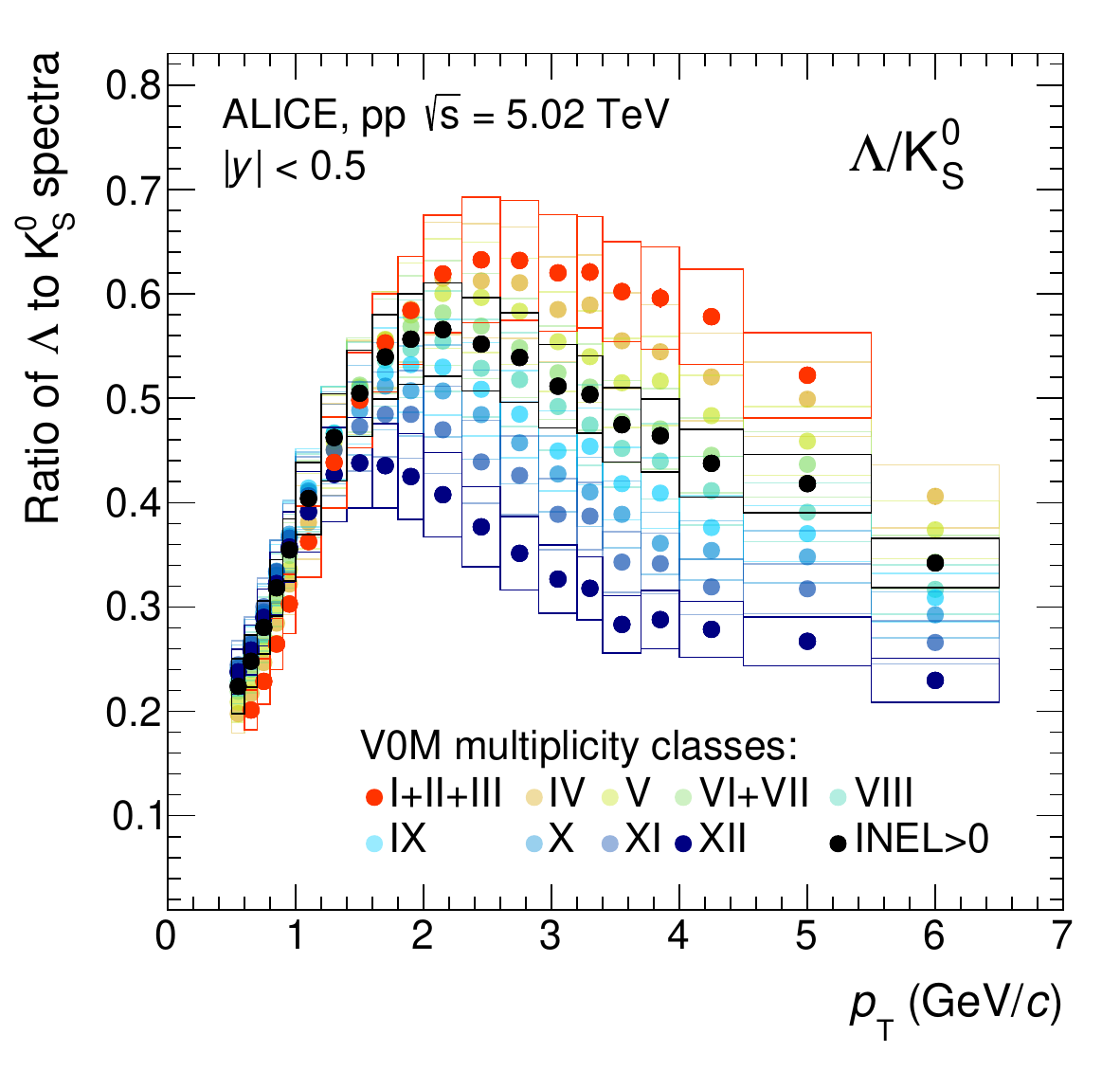}
\caption{\pt distribution of \lmb/\kzero ratio in pp collisions at \five and midrapidity for different multiplicity classes are shown with different colours. The lowest, highest and the \inelgz classes are highlighted.} 
\label{fig:lamovkzero}
\end{figure}

Another interesting observable that quantifies the multiplicity evolution of the \pt distributions is the ratio of the \lmb and \kzero spectra for each multiplicity class. This ratio is expected to evolve with multiplicity as the radial push experienced by baryons should be larger than that experienced by mesons if a collective medium is formed. Fig.~\ref{fig:lamovkzero} shows the \pt evolution of the \lmb/\kzero ratio for different multiplicity classes. Highlighted are the lowest, highest and \inelgz multiplicity classes. The ratio increases at low \pt, reaches a maximum between 1.5 and 3 \GeVc and then descends towards higher \pt. The position of the maximum evolves with multiplicity, being highest for the collisions with the highest multiplicity. The \inelgz ratio follows the lowest multiplicity trend at very low \pt, and then reaches higher values towards the highest multiplicity classes. This reflects the basic fact that the \inelgz spectra are dominated by low-multiplicity events at low \pt (where the spectra are softer) and by higher-multiplicity events at higher \pt (where the spectra are harder).

\begin{figure}[htb]
\centering
\includegraphics[scale=0.39]{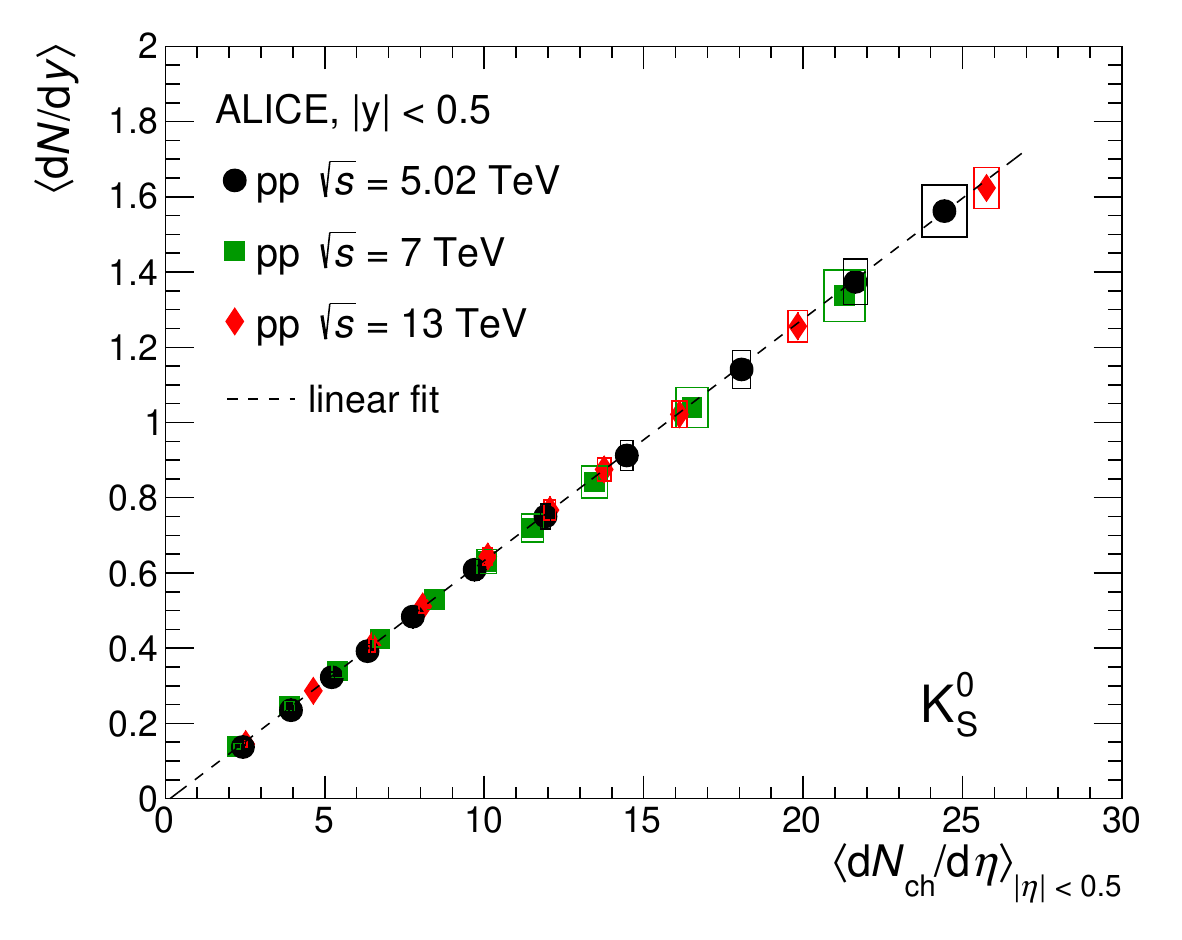}
\includegraphics[scale=0.39]{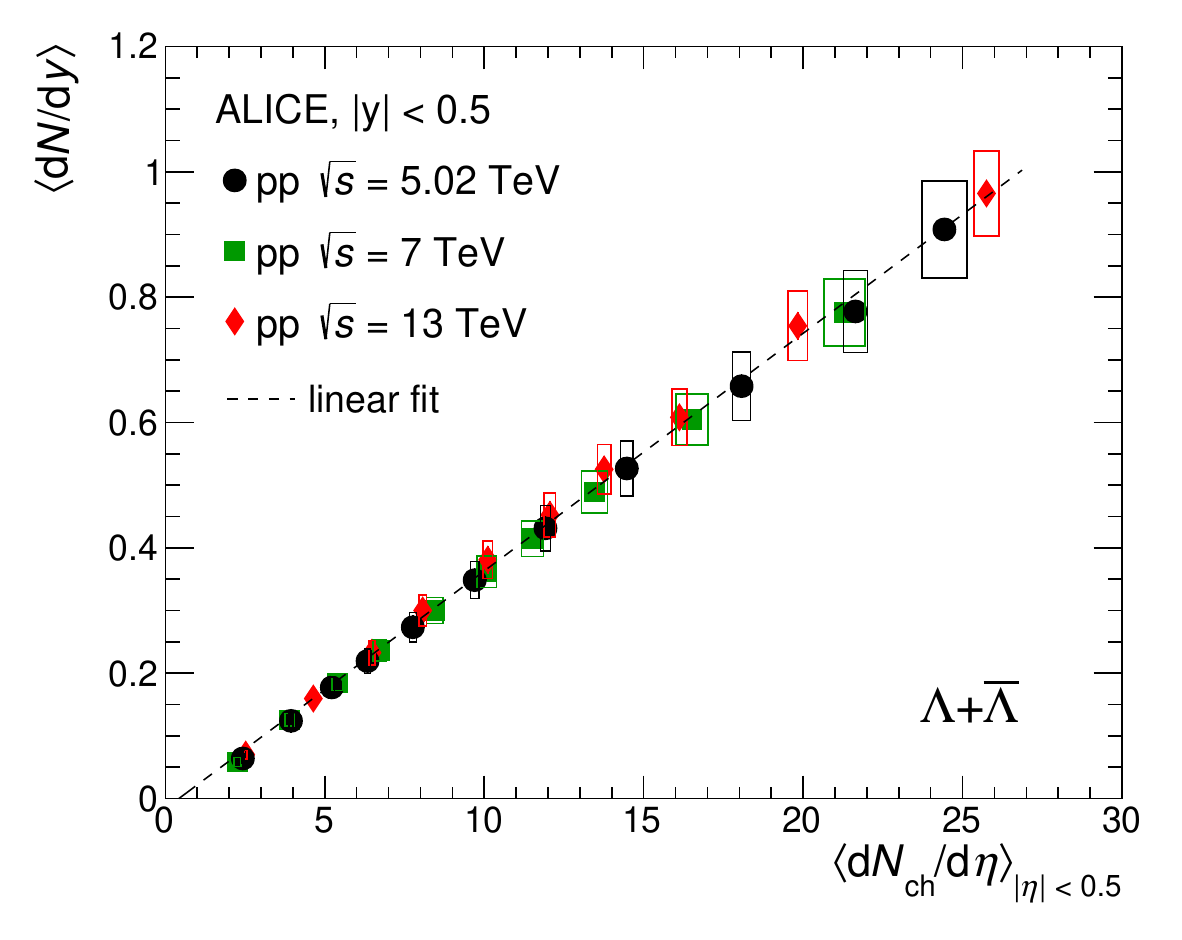}
\includegraphics[scale=0.39]{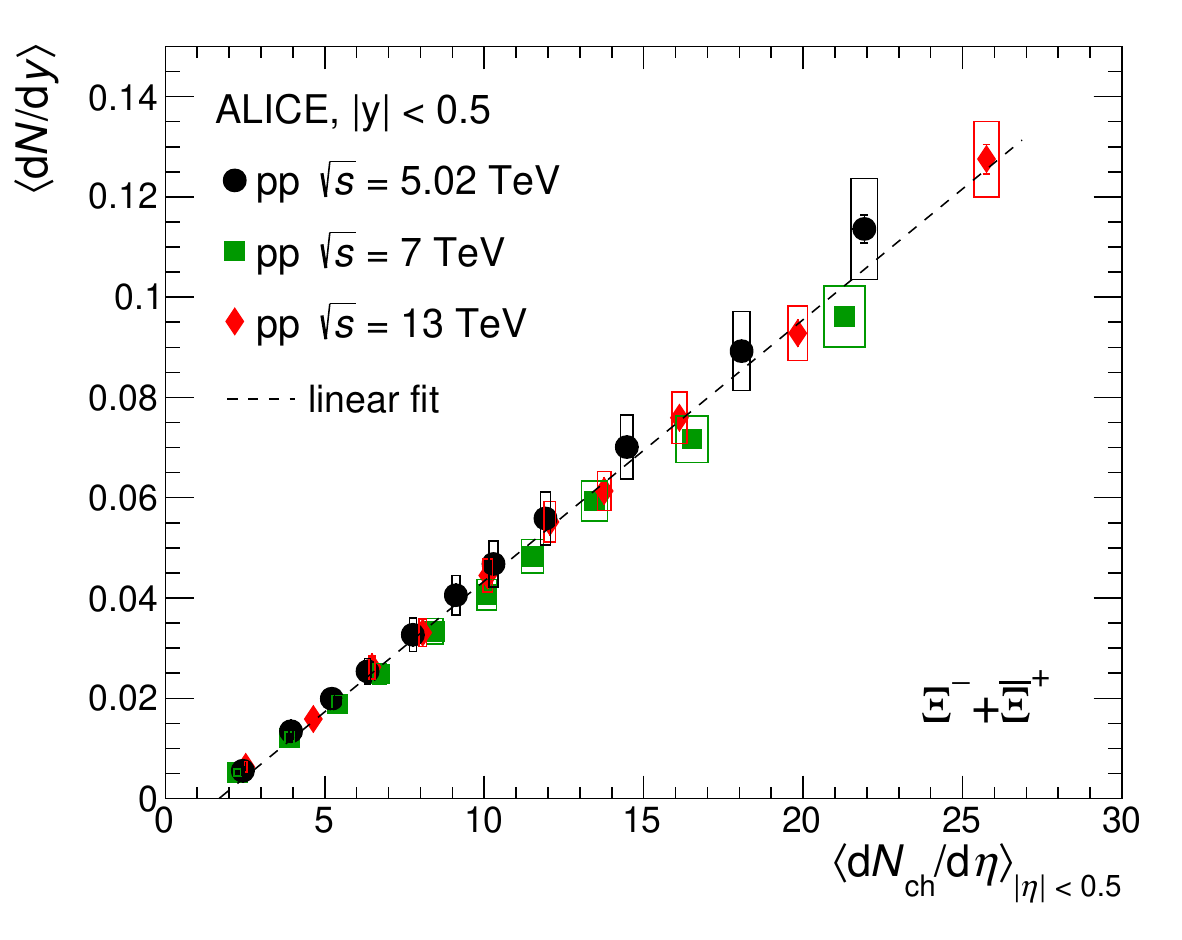}
\includegraphics[scale=0.39]{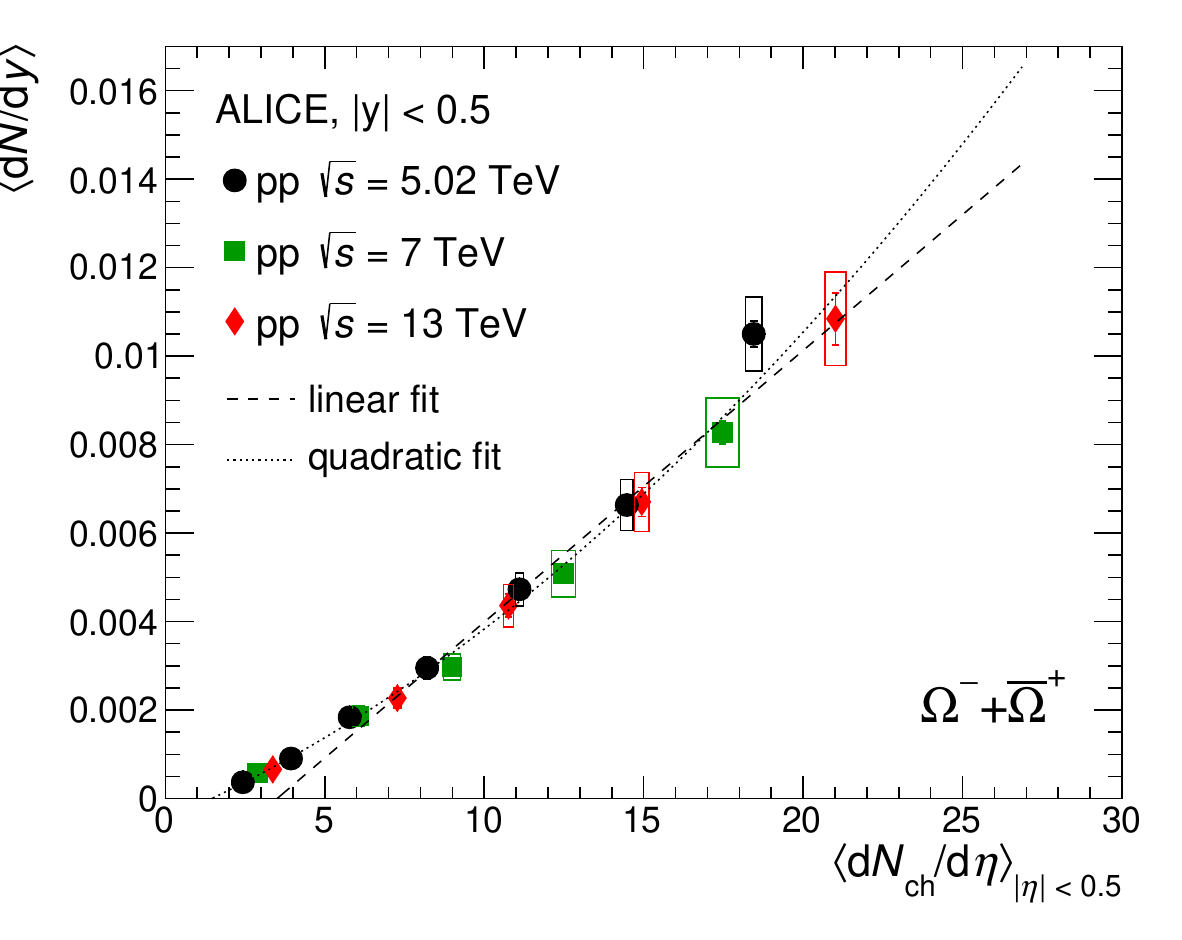}
\caption{\pt-integrated yields in pp collisions at different center-of-mass energies and midrapidity as a function of the event multiplicity for \kzero and \lmb (top panels), \Xiboth and \Omboth (bottom panels). \five results are shown as black dots, \seven results~\cite{ALICEstrPP7} as green squares and \thirteen results \cite{ALICEstrPP13} as red diamonds. Statistical uncertainties are represented as error bars and systematic uncertainties as boxes whose width represent the systematic error on the multiplicity determination. Linear combined fits are shown as dashed lines, while the quadratic fit to the \Omboth progression is shown as a dotted line.}
\label{fig:inetgratedyields}
\end{figure}

Integrated yields are presented in Fig.~\ref{fig:inetgratedyields}. The multiplicity progression is also compared to results at higher \pp energies, showing an excellent agreement of all three \pp energies. This demonstrates that the collision center-of-mass energy at the LHC does not affect strange particle production yields when similar multiplicities are at play.
It was proposed \cite{MorschLoizides} that the overall multiplicity evolution of the strange-particle over pion ratio may be fully determined by the minimum multiplicity necessary to produce one strange meson or baryon (pedestal effect). This interpretation starts from the observation that the d$N$/d$y$ dependence on multiplicity is approximately linear, which can now be tested with higher precision based on the present results. In Fig.~\ref{fig:inetgratedyields}, the dashed lines show the result of a linear combined fit to the experimental results at all center-of-mass energies, considering a quadratic sum of statistical and total systematic uncertainties. Systematic uncertainties are treated as uncorrelated across multiplicity, as a comprehensive assessment of their correlation with multiplicity is currently unavailable. A combined test of $\chi^2$ and Wald-Wolfowitz runs \cite{WWruns} shows that the trend is linear for \kzero, \lmb and \Xiboth ({\it p}-value $>$~0.3 in all cases). For the \Omboth baryon, instead of a linear fit ({\it p}-value $\sim 8\cdot 10^{-5}$), a quadratic trend describes the data better (dotted line, {\it p}-value $=~0.6$). Note that, despite its better reproduction of the \pp multiplicity ranges, the quadratic trend does not hold at multiplicities typical of \pPb and \PbPb interactions, as the trend of the ratio to pions would not saturate at the highest multiplicity, as was previously observed by ALICE \cite{ALICEstrPbPb276}.
The intercept of the linear fit with the $x$-axis (event multiplicity), the so-called ``pedestal'' ($P$) \cite{MorschLoizides}, was estimated for the different strange hadrons to be $P{\rm_{\kzero}} = 0.09 \pm 0.02~(\rm{stat.}) \pm 0.40~(syst.)$, $P_{\rm{\lmb}} = 0.30 \pm 0.03~(\rm{stat.}) \pm 0.72~(syst.)$, $P_{\rm{\Xiboth}} = 1.79 \pm 0.09~(\rm{stat.}) \pm 0.56~(syst.)$, $P_{\rm{\Omboth}} = 3.80 \pm 0.14~(\rm{stat.}) \pm 0.58~(syst.)$.

\section{Comparison to phenomenological models}
The reported results can be compared to state-of-the-art general-purpose Monte Carlo generators, such as PYTHIA, EPOS~4 and DCCI.\\
PYTHIA \cite{pythia8generic} is a phenomenological model incorporating pQCD processes and hadronization through colour string fragmentation~\cite{colstr}. The ``PYTHIA 8 Monash 2013'' tune, widely used in high-energy physics, implements multiple-parton interactions (MPI) and a basic version of the Color Reconnection (CR) mechanism \cite{pythia8cr}, which are both essential to reproduce the large variety of charged-particle multiplicity densities and \pt spectral shapes of unidentified hadrons measured in high-energy hadronic collisions. LHC results of identified particle production as a function of the multiplicity density necessitated the inclusion of more sophisticated CR mechanisms, implementing QCD-allowed 3-leg junctions \cite{ColRecBeyondLead}, and a modified string tension model in dense QCD environments, the so-called ``ropes'' mechanism \cite{bierlich2015effects}. This more recent version of the PYTHIA model will hereafter be called "PYTHIA 8 QCD-CR+Ropes".

EPOS~4 \cite{EPOS4general, EPOS4lf, EPOS4strang} is the newest version of the widely used EPOS model, based on the Gribov-Regge theory at parton level \cite{EPOSgrtheory}. Collective effects and modified hadronization in high-multiplicity environments are obtained by means of the ``core-corona'' mechanism \cite{EPOScoco}, which includes two different hadronization regimes: a high-density one (core), implementing statistical hadronization and hydrodynamic expansion, and an in-vacuum one (corona) which features string fragmentation. In the newest version of the model, some important phenomenological issues like energy conservation, factorization, and saturation are implemented in a consistent way.

The Dynamical Core-Corona Initialization model (DCCI) \cite{DCCIunif} is a state of the art implementation of the core-corona approach \cite{DCCIcoco}. In this case, the evolution of the system in its core and corona regions is dynamic, driven by the collisions that each parton experiences throughout the time evolution. Finally, hadronization can occur in vacuum (using PYTHIA) or using the Cooper-Frye formula from core droplets. 

In the models, multiplicity classes are defined as percentiles of the multiplicity distribution in the V0M geometrical acceptance region, adapting the class boundaries to match the corresponding \multdensmid values measured from the data. No extrapolation is necessary to extract yields and \avpt for the models, as their transverse momentum spectra extend to zero \pt.

\begin{figure}[]
\centering
\includegraphics[scale=0.36]{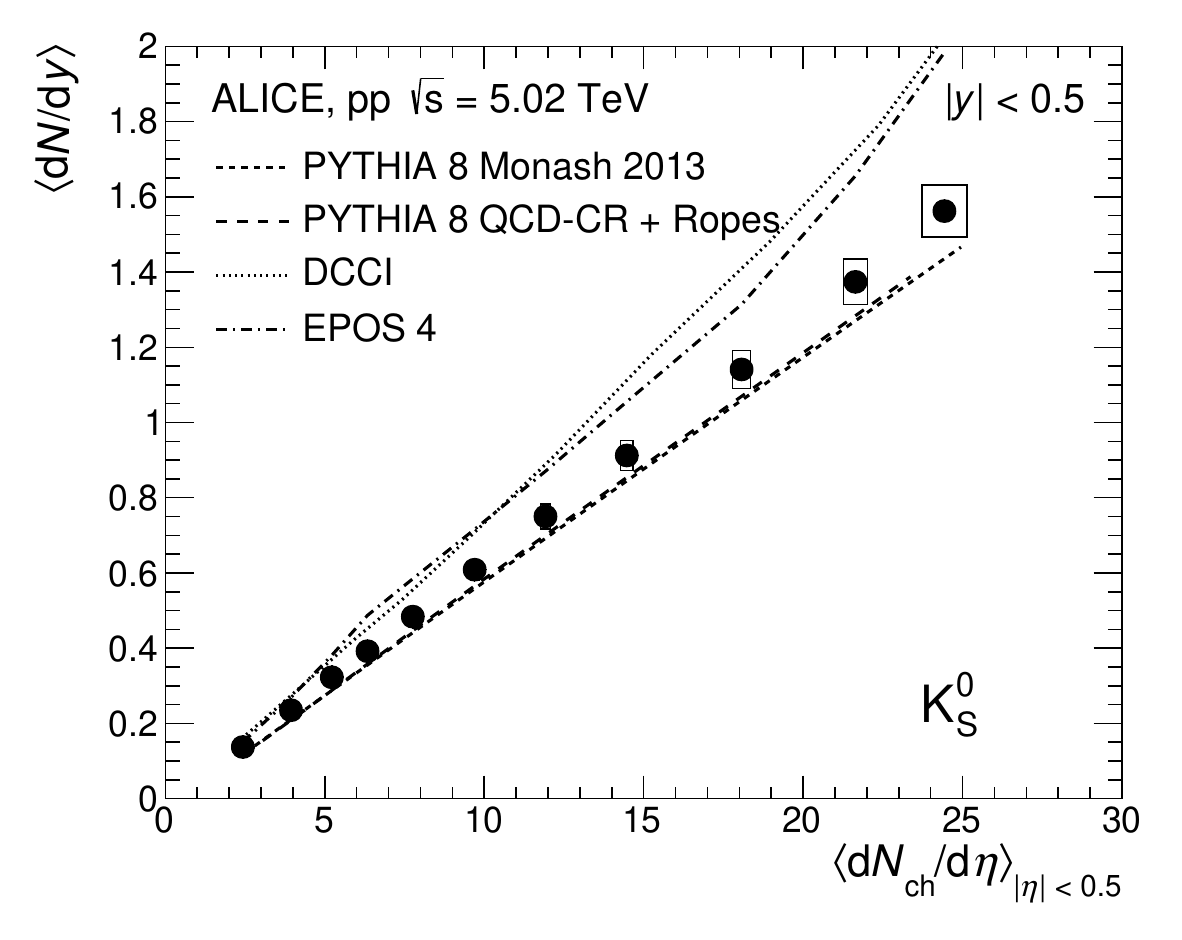}
\includegraphics[scale=0.36]{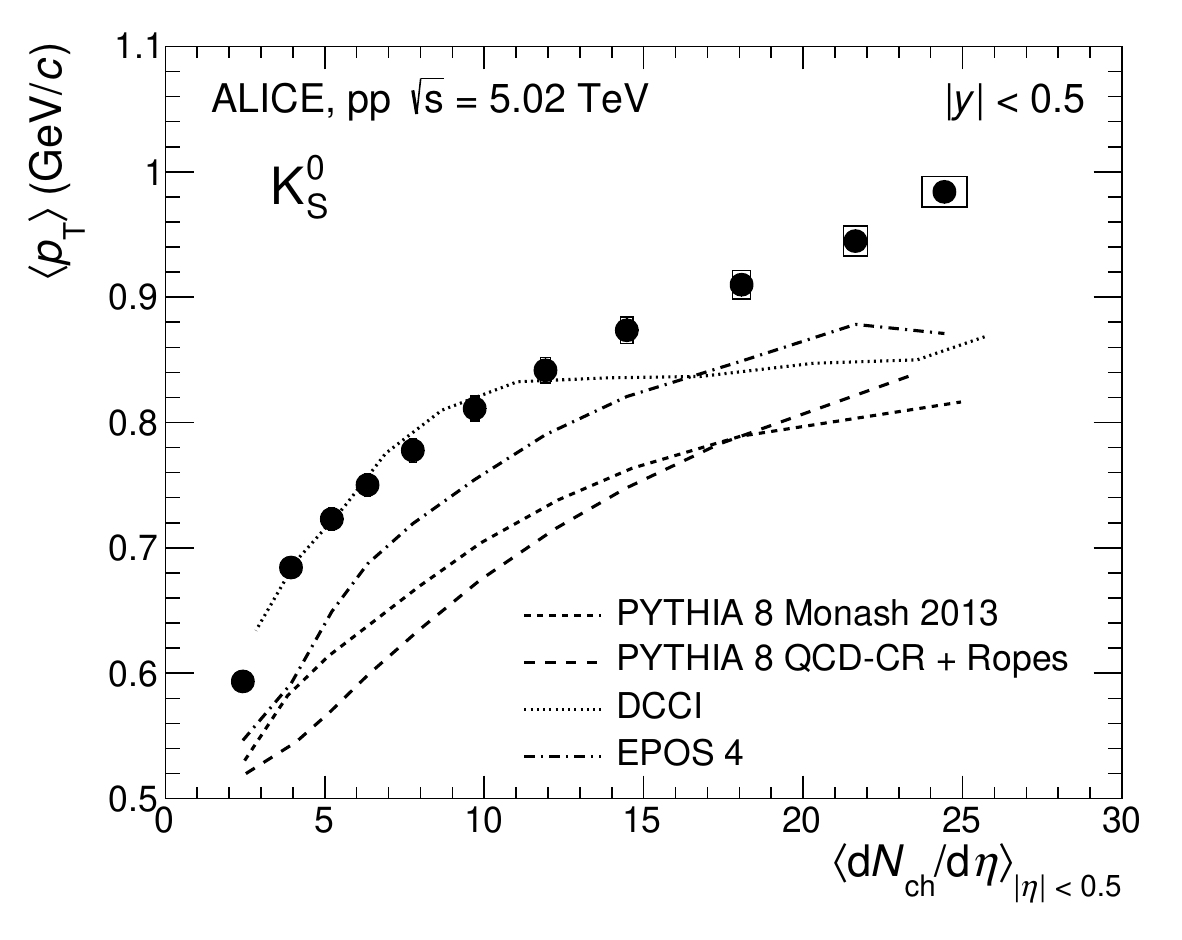}
\includegraphics[scale=0.36]{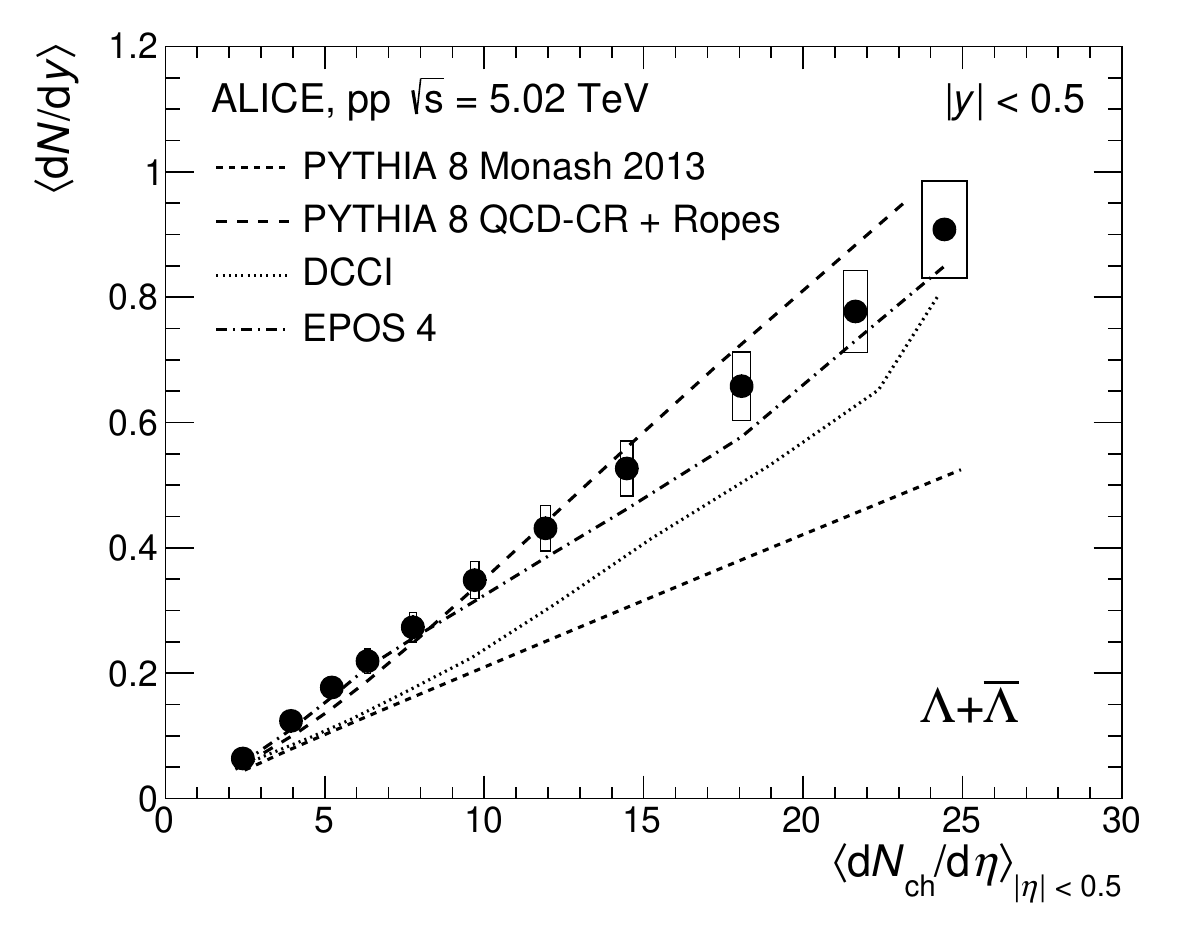}
\includegraphics[scale=0.36]{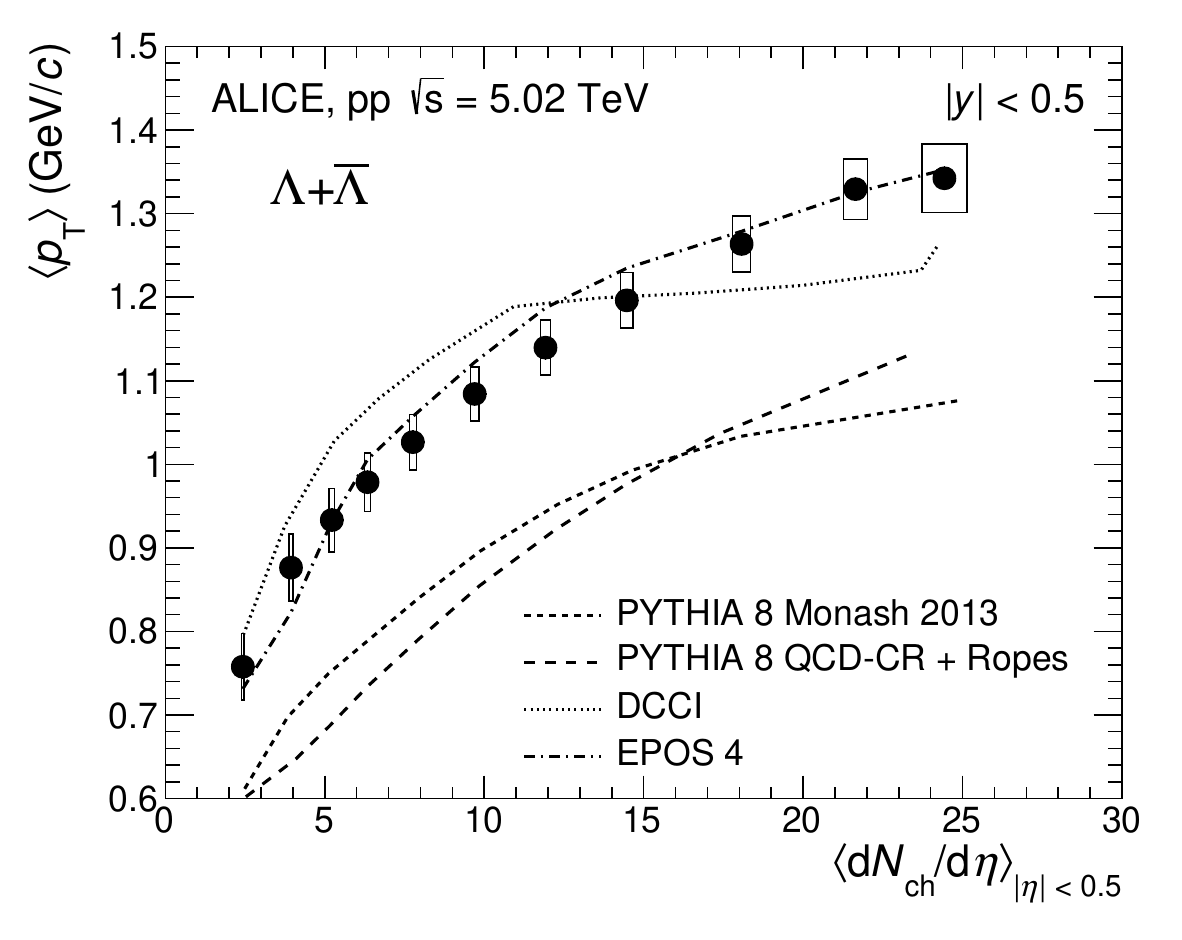}
\includegraphics[scale=0.36]{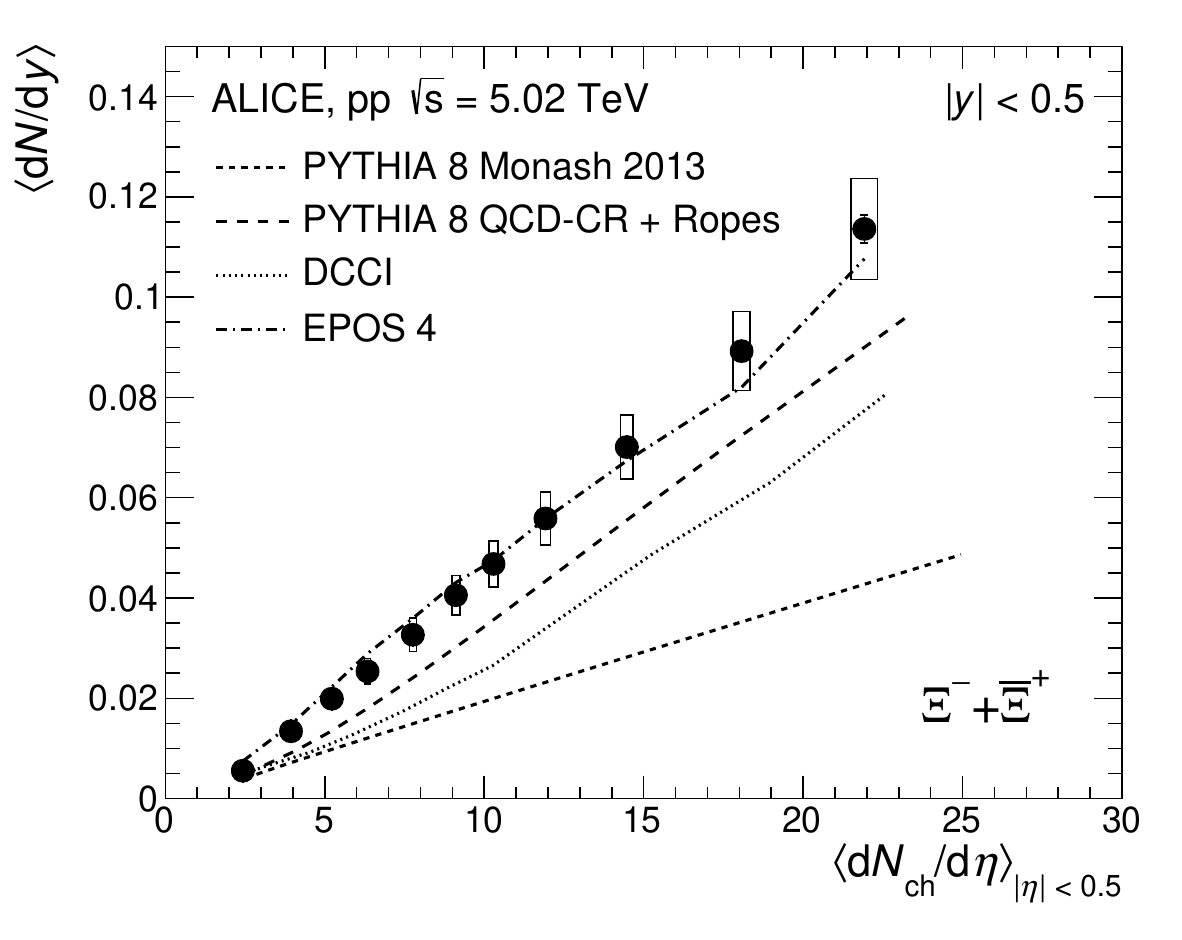}
\includegraphics[scale=0.36]{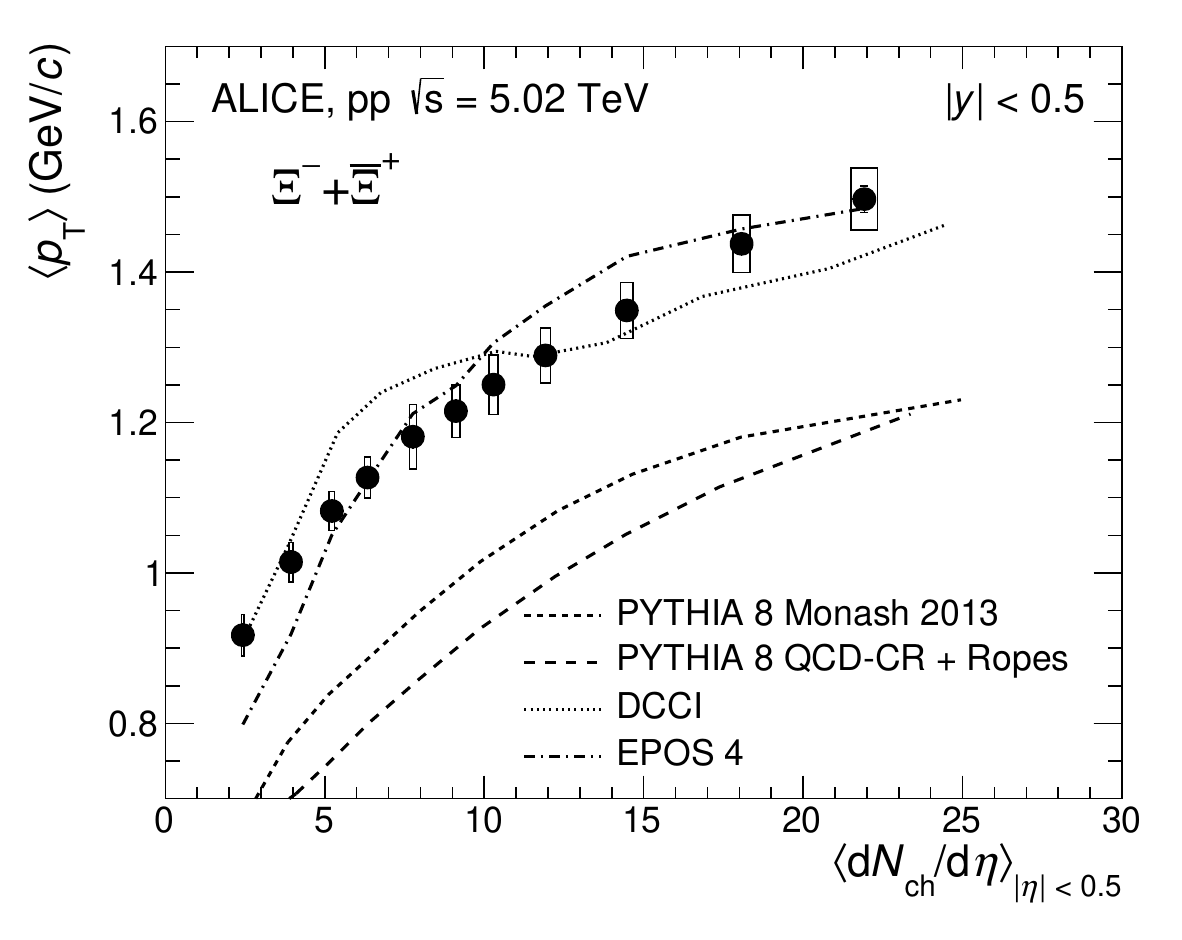}
\includegraphics[scale=0.36]{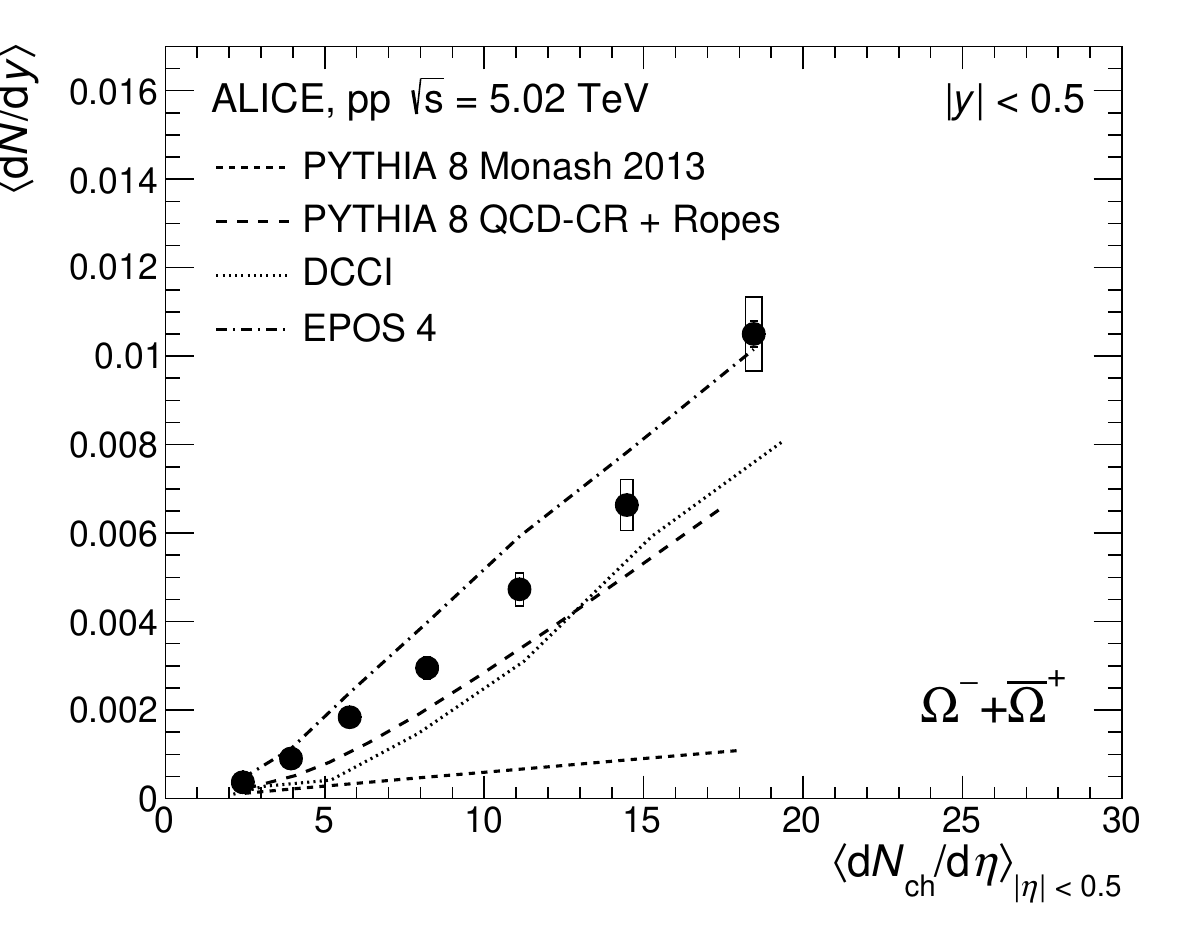}
\includegraphics[scale=0.36]{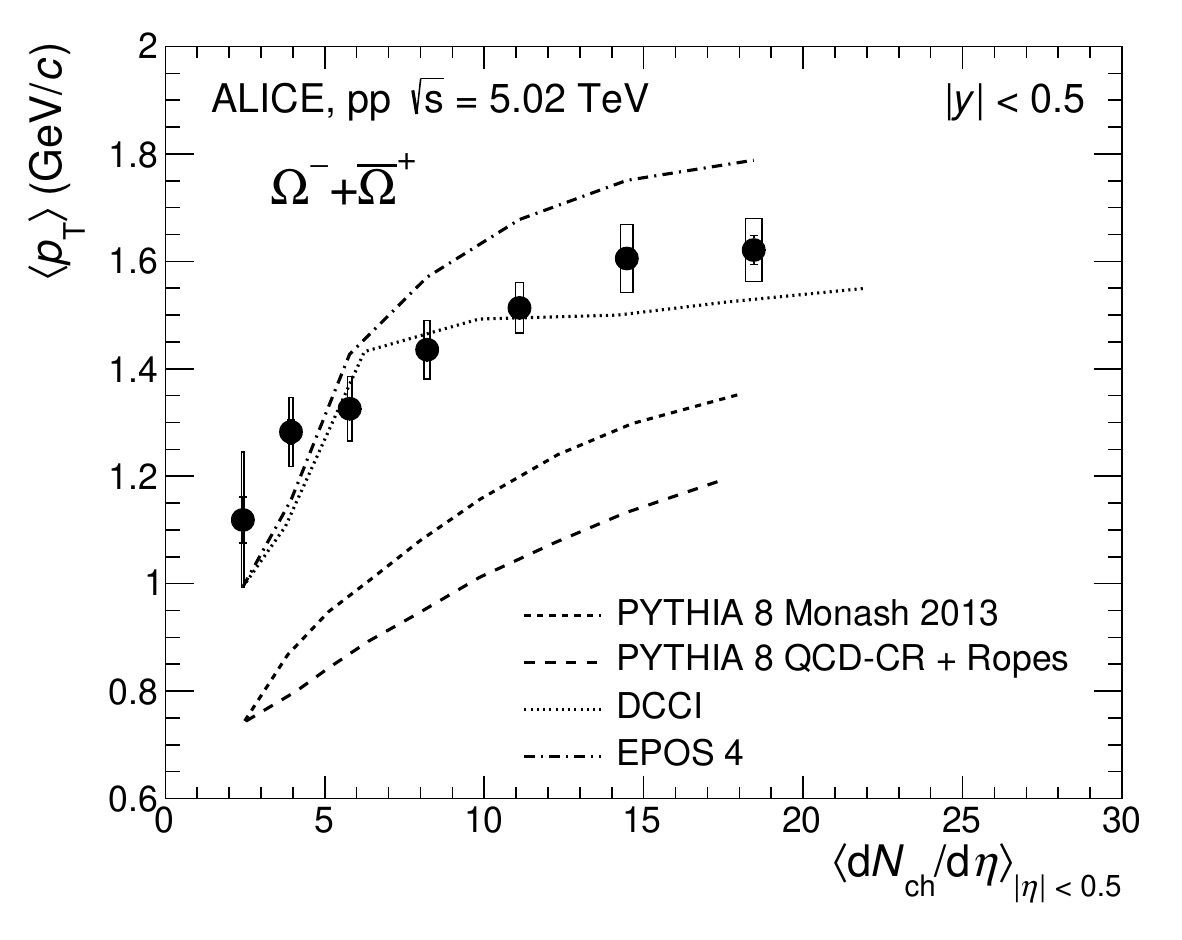}
\caption{\pt-integrated yields (left column) and \avpt (right column) at $\sqrt{s}$ = 5.02 TeV for \kzero, \lmb, \Xiboth, and \Omboth compared with MC generators PYTHIA 8 Monash 2013 (\cite{pythia8cr}, dashed line), PYTHIA 8 QCD-CR+Ropes (\cite{bierlich2015effects}, long-dashed line), DCCI (\cite{DCCIcoco}, dotted line) and EPOS~4 (\cite{EPOS4strang}, dash-dotted line).}
\label{fig:MccompYM}
\end{figure}

Figure~\ref{fig:MccompYM} reports the comparison of particle yields and \avpt to the aforementioned MC generators.
The improvement in the prediction of particle yields, when introducing QCD-inspired CR and the Ropes mechanism in PYTHIA, is noticeable for baryons, while the effect on mesons is negligible. The improvement, on the other hand, is absent with regards to the \avpt, where the increasing trend with multiplicity is better reproduced by the QCD-CR+Ropes tuning, but the spectra are overall softer in the model by about 20\%.
EPOS~4 provides a rather good description of the yields (slightly overshooting the \kzero and \Omboth trends) and a satisfactory description of \avpt for baryons (with some tension for \Omboth at high multiplicity), while for \kzero the EPOS~4 spectra are softer compared to the data across all multiplicities.
The DCCI model underpredicts the trend of the baryon yields with multiplicity, while slightly overshooting the one of \kzero. On the other hand, it describes the \avpt rather well for all particles, except for \kzero at very high multiplicity.
In general, although none of these models quantitatively describe both observables simultaneously for all strange hadrons, EPOS~4 provides the most accurate description for baryons. Note that the compatibility of the multiplicity trend of yields extracted at different $\sqrt{s}$ implies that the success of statistical model fits at higher energies~\cite{THERMUSsmallsys} would reflect in a very good comparison in the left column of Fig.~\ref{fig:MccompYM}

\begin{figure}[]
\centering
\hspace*{-1cm} 
\includegraphics[scale=0.83]{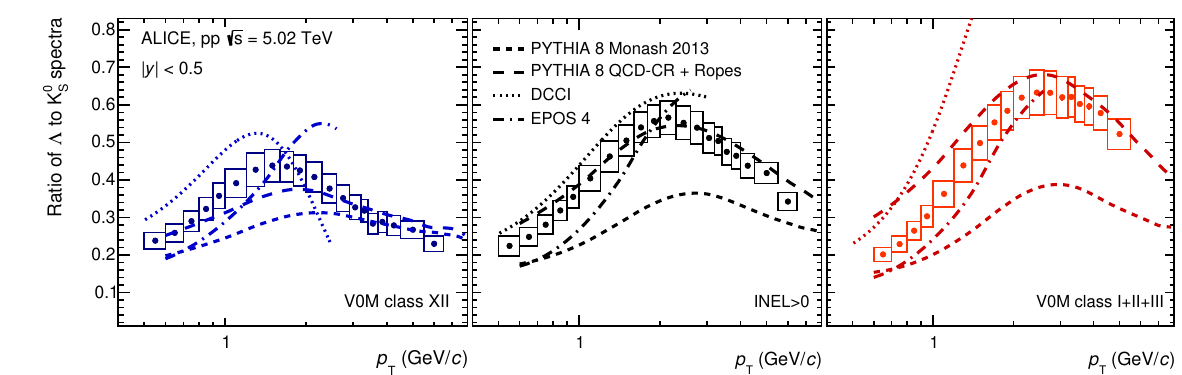}
\caption{\pt distribution of \lmb/\kzero ratio in the \inelgz event class (center) and low and high-multiplicity classes (left and right panels, respectively) from pp collisions at \five. Data are compared to PYTHIA 8 Monash 2013 (dashed lines), PYTHIA 8 QCD-QR+Ropes (long-dashed lines), DCCI (dotted lines) and EPOS~4 (dash-dotted lines) MC generators.}
\label{fig:lamovkzero_models}
\end{figure}

The \lmb/\kzero ratio is compared to the aforementioned MC generators in Fig. \ref{fig:lamovkzero_models} for the lowest and highest multiplicity classes (left and right panels) and for the \inelgz class (central panel).
Considering the PYTHIA 8 generator, the improvement introduced by the advanced CR model is again evident in predicting the maximum observed in the data at intermediate \pt. The Monash 2013 tune underpredicts all ratios and does not describe the shift of the peak with multiplicity. By contrast, the QCD-CR+Ropes tune improves the agreement with the data substantially, limiting the discrepancy to the very low \pt region, where the model under(over)predicts at low(high) multiplicity, while the agreement in the \inelgz case is satisfactory.
EPOS~4 has a reasonable agreement with the observed trend for \inelgz and at high multiplicity, but deviates at low multiplicity, where the model shows significant tensions both in the magnitude of the effect and in the position of the peak.
The DCCI model fails to predict the evolution of ratios with multiplicity: although the description of the \inelgz ratio is satisfactory, the low and high multiplicities show large incompatibilities.
Both the EPOS~4 and DCCI model curves terminate at 2.5 \GeVc to avoid large fluctuations due to limited statistics.

In general, all comparisons show that the data can be qualitatively described only if certain aspects of parton collectivity and modified hadronization in high-multiplicity collisions are considered and introduced in the models. Whether collectivity should be considered as parton flow (as in the two-component models EPOS~4 and DCCI) or Color Reconnection (as in PYTHIA) remains an open issue. Similarly, hadron yields can only be described if modified hadronization in a parton-dense environment is introduced: thermal hadronization is implemented in the case of EPOS~4 and DCCI, while Color Reconnection and Ropes account for the description of the observed yields in a model like PYTHIA.

\section{Conclusions}
The production of strange (\kzero and \lmb) and multi-strange (\Xiboth and \Omboth) hadrons was studied as a function of multiplicity in pp collisions at \five, exploiting a large data sample collected by the ALICE experiment during LHC Run 2.

Transverse momentum spectra, measured with unprecedented precision, show a distinctive hardening moving towards high multiplicity. The evolution stops at around 4~\GeVc, after which the ratio of the multiplicity selected to the \inelgz spectra approaches a saturation region.
The evolution of the spectra was quantitatively studied by extracting the \avpt and the \lmb/\kzero ratio. The \avpt increases with multiplicity, a feature that can be reproduced by phenomenological models only if certain aspects of collectivity at the partonic level are introduced. The \lmb/\kzero ratio features a structure with an intermediate-\pt maximum, which is reminiscent of the more prominent one observed in larger collision systems. Also in this case, models can qualitatively describe this ratio only if collectivity is introduced in the evolution of the system.

Particle yields were also studied and presented, extending previous works to lower energy (hence at slightly lower multiplicity) as well as improving the precision and granularity in the multiplicity selection, especially for multi-strange baryons.
Yields are observed to increase with multiplicity, in excellent agreement with what is observed at higher energies, demonstrating and confirming that the collision energy does not affect the average production probability of strange particles, whenever similar multiplicities are considered. The increase with multiplicity is consistent with a linear trend for \kzero, \lmb and \Xiboth, while a quadratic dependence better accommodates the \Omboth results. Such an increase is, however, not suited to describe the \Omboth yields at higher multiplicities, which are accessed in p--Pb and Pb--Pb collisions.
Models incorporating modified hadronization in high-density pp collisions qualitatively describe the data, whereas pure in-vacuum hadronization does not provide a good agreement.


\newenvironment{acknowledgement}{\relax}{\relax}
\begin{acknowledgement}
\section*{Acknowledgements}

The ALICE Collaboration would like to thank all its engineers and technicians for their invaluable contributions to the construction of the experiment and the CERN accelerator teams for the outstanding performance of the LHC complex.
The ALICE Collaboration gratefully acknowledges the resources and support provided by all Grid centres and the Worldwide LHC Computing Grid (WLCG) collaboration.
The ALICE Collaboration acknowledges the following funding agencies for their support in building and running the ALICE detector:
A. I. Alikhanyan National Science Laboratory (Yerevan Physics Institute) Foundation (ANSL), State Committee of Science and World Federation of Scientists (WFS), Armenia;
Austrian Academy of Sciences, Austrian Science Fund (FWF): [M 2467-N36] and Nationalstiftung f\"{u}r Forschung, Technologie und Entwicklung, Austria;
Ministry of Communications and High Technologies, National Nuclear Research Center, Azerbaijan;
Rede Nacional de Física de Altas Energias (Renafae), Financiadora de Estudos e Projetos (Finep), Funda\c{c}\~{a}o de Amparo \`{a} Pesquisa do Estado de S\~{a}o Paulo (FAPESP) and The Sao Paulo Research Foundation  (FAPESP), Brazil;
Bulgarian Ministry of Education and Science, within the National Roadmap for Research Infrastructures 2020-2027 (object CERN), Bulgaria;
Ministry of Education of China (MOEC) , Ministry of Science \& Technology of China (MSTC) and National Natural Science Foundation of China (NSFC), China;
Ministry of Science and Education and Croatian Science Foundation, Croatia;
Centro de Aplicaciones Tecnol\'{o}gicas y Desarrollo Nuclear (CEADEN), Cubaenerg\'{\i}a, Cuba;
Ministry of Education, Youth and Sports of the Czech Republic, Czech Republic;
The Danish Council for Independent Research | Natural Sciences, the VILLUM FONDEN and Danish National Research Foundation (DNRF), Denmark;
Helsinki Institute of Physics (HIP), Finland;
Commissariat \`{a} l'Energie Atomique (CEA) and Institut National de Physique Nucl\'{e}aire et de Physique des Particules (IN2P3) and Centre National de la Recherche Scientifique (CNRS), France;
Bundesministerium f\"{u}r Forschung, Technologie und Raumfahrt (BMFTR) and GSI Helmholtzzentrum f\"{u}r Schwerionenforschung GmbH, Germany;
General Secretariat for Research and Technology, Ministry of Education, Research and Religions, Greece;
National Research, Development and Innovation Office, Hungary;
Department of Atomic Energy Government of India (DAE), Department of Science and Technology, Government of India (DST), University Grants Commission, Government of India (UGC) and Council of Scientific and Industrial Research (CSIR), India;
National Research and Innovation Agency - BRIN, Indonesia;
Istituto Nazionale di Fisica Nucleare (INFN), Italy;
Japanese Ministry of Education, Culture, Sports, Science and Technology (MEXT) and Japan Society for the Promotion of Science (JSPS) KAKENHI, Japan;
Consejo Nacional de Ciencia (CONACYT) y Tecnolog\'{i}a, through Fondo de Cooperaci\'{o}n Internacional en Ciencia y Tecnolog\'{i}a (FONCICYT) and Direcci\'{o}n General de Asuntos del Personal Academico (DGAPA), Mexico;
Nederlandse Organisatie voor Wetenschappelijk Onderzoek (NWO), Netherlands;
The Research Council of Norway, Norway;
Pontificia Universidad Cat\'{o}lica del Per\'{u}, Peru;
Ministry of Science and Higher Education, National Science Centre and WUT ID-UB, Poland;
Korea Institute of Science and Technology Information and National Research Foundation of Korea (NRF), Republic of Korea;
Ministry of Education and Scientific Research, Institute of Atomic Physics, Ministry of Research and Innovation and Institute of Atomic Physics and Universitatea Nationala de Stiinta si Tehnologie Politehnica Bucuresti, Romania;
Ministerstvo skolstva, vyskumu, vyvoja a mladeze SR, Slovakia;
National Research Foundation of South Africa, South Africa;
Swedish Research Council (VR) and Knut \& Alice Wallenberg Foundation (KAW), Sweden;
European Organization for Nuclear Research, Switzerland;
Suranaree University of Technology (SUT), National Science and Technology Development Agency (NSTDA) and National Science, Research and Innovation Fund (NSRF via PMU-B B05F650021), Thailand;
Turkish Energy, Nuclear and Mineral Research Agency (TENMAK), Turkey;
National Academy of  Sciences of Ukraine, Ukraine;
Science and Technology Facilities Council (STFC), United Kingdom;
National Science Foundation of the United States of America (NSF) and United States Department of Energy, Office of Nuclear Physics (DOE NP), United States of America.
In addition, individual groups or members have received support from:
Czech Science Foundation (grant no. 23-07499S), Czech Republic;
FORTE project, reg.\ no.\ CZ.02.01.01/00/22\_008/0004632, Czech Republic, co-funded by the European Union, Czech Republic;
European Research Council (grant no. 950692), European Union;
Deutsche Forschungs Gemeinschaft (DFG, German Research Foundation) ``Neutrinos and Dark Matter in Astro- and Particle Physics'' (grant no. SFB 1258), Germany;
CONVECS project, CUP C97H23001700002 FESR 2021-2027 program, Italy;
FAIR - Future Artificial Intelligence Research, funded by the NextGenerationEU program (Italy).

\end{acknowledgement}

\section*{Data availability}
This manuscript has associated data in a HEPData repository at~\cite{hepdata.180658}.

\bibliographystyle{utphys}   
\bibliography{bibliography}
\newpage
\appendix

\section{The ALICE Collaboration}
\label{app:collab}
\begin{flushleft} 
\small

I.J.~Abualrob\,\orcidlink{0009-0005-3519-5631}\,$^{\rm 114}$, 
S.~Acharya\,\orcidlink{0000-0002-9213-5329}\,$^{\rm 50}$, 
G.~Aglieri Rinella\,\orcidlink{0000-0002-9611-3696}\,$^{\rm 32}$, 
L.~Aglietta\,\orcidlink{0009-0003-0763-6802}\,$^{\rm 24}$, 
N.~Agrawal\,\orcidlink{0000-0003-0348-9836}\,$^{\rm 25}$, 
Z.~Ahammed\,\orcidlink{0000-0001-5241-7412}\,$^{\rm 134}$, 
S.~Ahmad\,\orcidlink{0000-0003-0497-5705}\,$^{\rm 15}$, 
I.~Ahuja\,\orcidlink{0000-0002-4417-1392}\,$^{\rm 36}$, 
Z.~Akbar$^{\rm 81}$, 
A.~Akindinov\,\orcidlink{0000-0002-7388-3022}\,$^{\rm 140}$
% ...

\section*{Affiliation Notes}

$^{\rm I}$ Deceased\\
$^{\rm II}$ Also at: Max-Planck-Institut fur Physik, Munich, Germany\\
$^{\rm III}$ Also at: Czech Technical University in Prague, Prague, Czech Republic\\
$^{\rm IV}$ Also at: Instituto de Fisica da Universidade de Sao Paulo\\
$^{\rm V}$ Also at: Dipartimento DET del Politecnico di Torino, Turin, Italy\\
$^{\rm VI}$ Also at: University College of Dublin, Dublin, Ireland\\
$^{\rm VII}$ Also at: Department of Applied Physics, Aligarh Muslim University, Aligarh, India\\
$^{\rm VIII}$ Also at: Institute of Theoretical Physics, University of Wroclaw, Wroclaw, Poland\\
$^{\rm IX}$ Also at: Facultad de Ciencias, Universidad Nacional Aut\'{o}noma de M\'{e}xico, Mexico City, Mexico\\

\section*{Collaboration Institutes}

$^{1}$ A.I. Alikhanyan National Science Laboratory (Yerevan Physics Institute) Foundation, Yerevan, Armenia\\
$^{2}$ AGH University of Krakow, Cracow, Poland\\
$^{3}$ Bogolyubov Institute for Theoretical Physics, National Academy of Sciences of Ukraine, Kyiv, Ukraine\\
$^{4}$ Bose Institute, Department of Physics  and Centre for Astroparticle Physics and Space Science (CAPSS), Kolkata, India\\
$^{5}$ California Polytechnic State University, San Luis Obispo, California, United States\\
$^{6}$ Central China Normal University, Wuhan, China\\
$^{7}$ Centro de Aplicaciones Tecnol\'{o}gicas y Desarrollo Nuclear (CEADEN), Havana, Cuba\\
$^{8}$ Centro de Investigaci\'{o}n y de Estudios Avanzados (CINVESTAV), Mexico City and M\'{e}rida, Mexico\\
$^{9}$ Chicago State University, Chicago, Illinois, United States\\
$^{10}$ China Nuclear Data Center, China Institute of Atomic Energy, Beijing, China\\
$^{11}$ China University of Geosciences, Wuhan, China\\
$^{12}$ Chungbuk National University, Cheongju, Republic of Korea\\
$^{13}$ Comenius University Bratislava, Faculty of Mathematics, Physics and Informatics, Bratislava, Slovak Republic\\
$^{14}$ Creighton University, Omaha, Nebraska, United States\\
$^{15}$ Department of Physics, Aligarh Muslim University, Aligarh, India\\
$^{16}$ Department of Physics, Pusan National University, Pusan, Republic of Korea\\
$^{17}$ Department of Physics, Sejong University, Seoul, Republic of Korea\\
$^{18}$ Department of Physics, University of California, Berkeley, California, United States\\
$^{19}$ Department of Physics, University of Oslo, Oslo, Norway\\
$^{20}$ Department of Physics and Technology, University of Bergen, Bergen, Norway\\
$^{21}$ Dipartimento di Fisica, Universit\`{a} di Pavia, Pavia, Italy\\
$^{22}$ Dipartimento di Fisica dell'Universit\`{a} and Sezione INFN, Cagliari, Italy\\
$^{23}$ Dipartimento di Fisica dell'Universit\`{a} and Sezione INFN, Trieste, Italy\\
$^{24}$ Dipartimento di Fisica dell'Universit\`{a} and Sezione INFN, Turin, Italy\\
$^{25}$ Dipartimento di Fisica e Astronomia dell'Universit\`{a} and Sezione INFN, Bologna, Italy\\
$^{26}$ Dipartimento di Fisica e Astronomia dell'Universit\`{a} and Sezione INFN, Catania, Italy\\
$^{27}$ Dipartimento di Fisica e Astronomia dell'Universit\`{a} and Sezione INFN, Padova, Italy\\
$^{28}$ Dipartimento di Fisica `E.R.~Caianiello' dell'Universit\`{a} and Gruppo Collegato INFN, Salerno, Italy\\
$^{29}$ Dipartimento DISAT del Politecnico and Sezione INFN, Turin, Italy\\
$^{30}$ Dipartimento di Scienze MIFT, Universit\`{a} di Messina, Messina, Italy\\
$^{31}$ Dipartimento Interateneo di Fisica `M.~Merlin' and Sezione INFN, Bari, Italy\\
$^{32}$ European Organization for Nuclear Research (CERN), Geneva, Switzerland\\
$^{33}$ Faculty of Electrical Engineering, Mechanical Engineering and Naval Architecture, University of Split, Split, Croatia\\
$^{34}$ Faculty of Nuclear Sciences and Physical Engineering, Czech Technical University in Prague, Prague, Czech Republic\\
$^{35}$ Faculty of Physics, Sofia University, Sofia, Bulgaria\\
$^{36}$ Faculty of Science, P.J.~\v{S}af\'{a}rik University, Ko\v{s}ice, Slovak Republic\\
$^{37}$ Faculty of Technology, Environmental and Social Sciences, Bergen, Norway\\
$^{38}$ Frankfurt Institute for Advanced Studies, Johann Wolfgang Goethe-Universit\"{a}t Frankfurt, Frankfurt, Germany\\
$^{39}$ Fudan University, Shanghai, China\\
$^{40}$ Gangneung-Wonju National University, Gangneung, Republic of Korea\\
$^{41}$ Gauhati University, Department of Physics, Guwahati, India\\
$^{42}$ Helmholtz-Institut f\"{u}r Strahlen- und Kernphysik, Rheinische Friedrich-Wilhelms-Universit\"{a}t Bonn, Bonn, Germany\\
$^{43}$ Helsinki Institute of Physics (HIP), Helsinki, Finland\\
$^{44}$ High Energy Physics Group,  Universidad Aut\'{o}noma de Puebla, Puebla, Mexico\\
$^{45}$ Horia Hulubei National Institute of Physics and Nuclear Engineering, Bucharest, Romania\\
$^{46}$ HUN-REN Wigner Research Centre for Physics, Budapest, Hungary\\
$^{47}$ Indian Institute of Technology Bombay (IIT), Mumbai, India\\
$^{48}$ Indian Institute of Technology Indore, Indore, India\\
$^{49}$ INFN, Laboratori Nazionali di Frascati, Frascati, Italy\\
$^{50}$ INFN, Sezione di Bari, Bari, Italy\\
$^{51}$ INFN, Sezione di Bologna, Bologna, Italy\\
$^{52}$ INFN, Sezione di Cagliari, Cagliari, Italy\\
$^{53}$ INFN, Sezione di Catania, Catania, Italy\\
$^{54}$ INFN, Sezione di Padova, Padova, Italy\\
$^{55}$ INFN, Sezione di Pavia, Pavia, Italy\\
$^{56}$ INFN, Sezione di Torino, Turin, Italy\\
$^{57}$ INFN, Sezione di Trieste, Trieste, Italy\\
$^{58}$ Inha University, Incheon, Republic of Korea\\
$^{59}$ Institute for Gravitational and Subatomic Physics (GRASP), Utrecht University/Nikhef, Utrecht, Netherlands\\
$^{60}$ Institute of Experimental Physics, Slovak Academy of Sciences, Ko\v{s}ice, Slovak Republic\\
$^{61}$ Institute of Physics, Homi Bhabha National Institute, Bhubaneswar, India\\
$^{62}$ Institute of Physics of the Czech Academy of Sciences, Prague, Czech Republic\\
$^{63}$ Institute of Space Science (ISS), Bucharest, Romania\\
$^{64}$ Institut f\"{u}r Kernphysik, Johann Wolfgang Goethe-Universit\"{a}t Frankfurt, Frankfurt, Germany\\
$^{65}$ Instituto de Ciencias Nucleares, Universidad Nacional Aut\'{o}noma de M\'{e}xico, Mexico City, Mexico\\
$^{66}$ Instituto de F\'{i}sica, Universidade Federal do Rio Grande do Sul (UFRGS), Porto Alegre, Brazil\\
$^{67}$ Instituto de F\'{\i}sica, Universidad Nacional Aut\'{o}noma de M\'{e}xico, Mexico City, Mexico\\
$^{68}$ iThemba LABS, National Research Foundation, Somerset West, South Africa\\
$^{69}$ Jeonbuk National University, Jeonju, Republic of Korea\\
$^{70}$ Korea Institute of Science and Technology Information, Daejeon, Republic of Korea\\
$^{71}$ Laboratoire de Physique Subatomique et de Cosmologie, Universit\'{e} Grenoble-Alpes, CNRS-IN2P3, Grenoble, France\\
$^{72}$ Lawrence Berkeley National Laboratory, Berkeley, California, United States\\
$^{73}$ Lund University Department of Physics, Division of Particle Physics, Lund, Sweden\\
$^{74}$ Marietta Blau Institute, Vienna, Austria\\
$^{75}$ Nagasaki Institute of Applied Science, Nagasaki, Japan\\
$^{76}$ Nara Women{'}s University (NWU), Nara, Japan\\
$^{77}$ National and Kapodistrian University of Athens, School of Science, Department of Physics , Athens, Greece\\
$^{78}$ National Centre for Nuclear Research, Warsaw, Poland\\
$^{79}$ National Institute of Science Education and Research, Homi Bhabha National Institute, Jatni, India\\
$^{80}$ National Nuclear Research Center, Baku, Azerbaijan\\
$^{81}$ National Research and Innovation Agency - BRIN, Jakarta, Indonesia\\
$^{82}$ Niels Bohr Institute, University of Copenhagen, Copenhagen, Denmark\\
$^{83}$ Nikhef, National institute for subatomic physics, Amsterdam, Netherlands\\
$^{84}$ Nuclear Physics Group, STFC Daresbury Laboratory, Daresbury, United Kingdom\\
$^{85}$ Nuclear Physics Institute of the Czech Academy of Sciences, Husinec-\v{R}e\v{z}, Czech Republic\\
$^{86}$ Oak Ridge National Laboratory, Oak Ridge, Tennessee, United States\\
$^{87}$ Ohio State University, Columbus, Ohio, United States\\
$^{88}$ Physics department, Faculty of science, University of Zagreb, Zagreb, Croatia\\
$^{89}$ Physics Department, Panjab University, Chandigarh, India\\
$^{90}$ Physics Department, University of Jammu, Jammu, India\\
$^{91}$ Physics Program and International Institute for Sustainability with Knotted Chiral Meta Matter (WPI-SKCM$^{2}$), Hiroshima University, Hiroshima, Japan\\
$^{92}$ Physikalisches Institut, Eberhard-Karls-Universit\"{a}t T\"{u}bingen, T\"{u}bingen, Germany\\
$^{93}$ Physikalisches Institut, Ruprecht-Karls-Universit\"{a}t Heidelberg, Heidelberg, Germany\\
$^{94}$ Physik Department, Technische Universit\"{a}t M\"{u}nchen, Munich, Germany\\
$^{95}$ Politecnico di Bari and Sezione INFN, Bari, Italy\\
$^{96}$ Research Division and ExtreMe Matter Institute EMMI, GSI Helmholtzzentrum f\"ur Schwerionenforschung GmbH, Darmstadt, Germany\\
$^{97}$ Saga University, Saga, Japan\\
$^{98}$ Saha Institute of Nuclear Physics, Homi Bhabha National Institute, Kolkata, India\\
$^{99}$ School of Physics and Astronomy, University of Birmingham, Birmingham, United Kingdom\\
$^{100}$ Secci\'{o}n F\'{\i}sica, Departamento de Ciencias, Pontificia Universidad Cat\'{o}lica del Per\'{u}, Lima, Peru\\
$^{101}$ SUBATECH, IMT Atlantique, Nantes Universit\'{e}, CNRS-IN2P3, Nantes, France\\
$^{102}$ Sungkyunkwan University, Suwon City, Republic of Korea\\
$^{103}$ Suranaree University of Technology, Nakhon Ratchasima, Thailand\\
$^{104}$ Technical University of Ko\v{s}ice, Ko\v{s}ice, Slovak Republic\\
$^{105}$ The Henryk Niewodniczanski Institute of Nuclear Physics, Polish Academy of Sciences, Cracow, Poland\\
$^{106}$ The University of Texas at Austin, Austin, Texas, United States\\
$^{107}$ Universidad Aut\'{o}noma de Sinaloa, Culiac\'{a}n, Mexico\\
$^{108}$ Universidade de S\~{a}o Paulo (USP), S\~{a}o Paulo, Brazil\\
$^{109}$ Universidade Estadual de Campinas (UNICAMP), Campinas, Brazil\\
$^{110}$ Universidade Federal do ABC, Santo Andre, Brazil\\
$^{111}$ Universitatea Nationala de Stiinta si Tehnologie Politehnica Bucuresti, Bucharest, Romania\\
$^{112}$ University of Cape Town, Cape Town, South Africa\\
$^{113}$ University of Derby, Derby, United Kingdom\\
$^{114}$ University of Houston, Houston, Texas, United States\\
$^{115}$ University of Jyv\"{a}skyl\"{a}, Jyv\"{a}skyl\"{a}, Finland\\
$^{116}$ University of Kansas, Lawrence, Kansas, United States\\
$^{117}$ University of Liverpool, Liverpool, United Kingdom\\
$^{118}$ University of Science and Technology of China, Hefei, China\\
$^{119}$ University of Silesia in Katowice, Katowice, Poland\\
$^{120}$ University of South-Eastern Norway, Kongsberg, Norway\\
$^{121}$ University of Tennessee, Knoxville, Tennessee, United States\\
$^{122}$ University of the Witwatersrand, Johannesburg, South Africa\\
$^{123}$ University of Tokyo, Tokyo, Japan\\
$^{124}$ University of Tsukuba, Tsukuba, Japan\\
$^{125}$ Universit\"{a}t M\"{u}nster, Institut f\"{u}r Kernphysik, M\"{u}nster, Germany\\
$^{126}$ Universit\'{e} Clermont Auvergne, CNRS/IN2P3, LPC, Clermont-Ferrand, France\\
$^{127}$ Universit\'{e} de Lyon, CNRS/IN2P3, Institut de Physique des 2 Infinis de Lyon, Lyon, France\\
$^{128}$ Universit\'{e} de Strasbourg, CNRS, IPHC UMR 7178, F-67000 Strasbourg, France, Strasbourg, France\\
$^{129}$ Universit\'{e} Paris-Saclay, Centre d'Etudes de Saclay (CEA), IRFU, D\'{e}partment de Physique Nucl\'{e}aire (DPhN), Saclay, France\\
$^{130}$ Universit\'{e}  Paris-Saclay, CNRS/IN2P3, IJCLab, Orsay, France\\
$^{131}$ Universit\`{a} degli Studi di Foggia, Foggia, Italy\\
$^{132}$ Universit\`{a} del Piemonte Orientale, Vercelli, Italy\\
$^{133}$ Universit\`{a} di Brescia, Brescia, Italy\\
$^{134}$ Variable Energy Cyclotron Centre, Homi Bhabha National Institute, Kolkata, India\\
$^{135}$ Warsaw University of Technology, Warsaw, Poland\\
$^{136}$ Wayne State University, Detroit, Michigan, United States\\
$^{137}$ Yale University, New Haven, Connecticut, United States\\
$^{138}$ Yildiz Technical University, Istanbul, Turkey\\
$^{139}$ Yonsei University, Seoul, Republic of Korea\\
$^{140}$ Affiliated with an institute formerly covered by a cooperation agreement with CERN\\
$^{141}$ Affiliated with an international laboratory covered by a cooperation agreement with CERN.\\

\end{flushleft} 

\end{document}